\documentclass[useAMS,usenatbib,usegraphicx]{mn2e}
\usepackage{times}

\def\la{\raise.5ex\hbox{$<$}\kern-.8em\lower 1mm\hbox{$\sim$}}
\def\ma{\raise.5ex\hbox{$>$}\kern-.8em\lower 1mm\hbox{$\sim$}}

\def\kms{$\rm km\, s^{-1}$}
\def\cm3{$\rm cm^{-3}$}

\def\Vs{$\rm V_{s}$}
\def\n0{$\rm n_{0}$}
\def\B0{$\rm B_{0}$}

\def\Fh{$\rm F_{h}$~}
\def\erg{$\rm erg\, cm^{-2}\, s^{-1}$}
\def\mum{$\mu$m}

\def\L12{L$_{12\mu m}$~}
\def\F12{F$_{12\mu m}$~}
\def\agr{a$_{gr}$}
\def\Hb{H$\beta$}

\title[The narrow line  region of  Mrk 766]{The continuum and narrow line   
 region of the NLS1 galaxy Mrk 766 }
\author[Rodr\'{\i}guez-Ardila et al.]{ A. Rodr\'{\i}guez-Ardila$^1$\thanks{Visiting Astronomer at the Infrared Telescope Facility, which is operated by the University of Hawaii under Cooperative Agreement no. NCC 5-538 with the National Aeronautics and Space Administration, Office of Space Science, Planetary Astronomy Program.}, M. Contini$^{2}$, S. M. Viegas$^{3}$ \\
$^1$ Laborat\'{o}rio Nacional de Astrof\'{i}sica - Rua dos Estados Unidos 154,
Bairro das Nac\~oes.
CEP 37504-364, Itajub\'{a}, MG, Brazil\\
$^{2}$School of Physics and Astronomy, Tel Aviv University, Tel Aviv
69978, Israel \\
$^{3}$Instituto de Astronomia, Geof\'{\i}sica e Ci\^encias
Atmosf\'ericas - USP, Rua do Mat\~ao 1226, 05508-900 São Paulo SP, Brazil\\
}

\begin{document}

\date{Accepted in MNRAS}

\pagerange{\pageref{firstpage}--\pageref{lastpage}} \pubyear{2003}

\maketitle

\label{firstpage}

\begin{abstract}

We present the first spectroscopic observations in the interval
0.8$\mu$m$-$4.0$\mu$m, 
complemented with existing {\it HST} UV and optical spectroscopy, of the
narrow-line Seyfert 1 galaxy Mrk 766. The NIR spectrum is 
characterized by numerous
permitted lines of H\,{\sc i}, He\,{\sc i}, He\,{\sc ii} and 
Fe\,{\sc ii}, and forbidden lines of [S\,{\sc ii}], [S\,{\sc iii}]
and [Fe\,{\sc ii}] among others. High
ionized species such as [Si\,{\sc ix}], [Si\,{\sc x}],
[S\,{\sc ix}] and [Mg\,{\sc vii}] are also observed. The continuum emission
has a complex shape, with contribution of the central engine,
circumnuclear stellar population and dust. This last component is
evidenced by the presence of an excess of emission peaking at
2.25$\mu$m, well fitted by blackbody function with $T_{\rm bb}$=1200~K.
That temperature is close to the evaporation temperature of graphite
grains. As such, it provides strong evidence of hot dust, probably
very close to the nucleus. Consistent modeling of 
the line spectrum and the broad band continuum  
by composite models, which account  for the photoionizing flux  of
the central engine and shocks, shows that shock velocities range between 100 and 
500 \kms, the preshock densities between 100 and 1000 \cm3 and the 
radiation fluxes from the active centre  between 10$^9$ and 5 10$^{12}$ 
photons cm$^{-2}$ s$^{-1}$ eV$^{-1}$ at 1 Ryd with spectral indices  
$\alpha_{UV}$ = -1.5 and $\alpha_X$=-0.4.
Adopting silicate grains, dust-to-gas ratios are between 10$^{-6}$ and 
4 10$^{-4}$ by mass. The emitting clouds are at an average distance of 
160~pc from the centre, with high velocity clouds closer and low 
velocity clouds farther from the centre. The N/H relative abundance 
deduced from the fit of the [N\,{\sc ii}] 6548+/[O\,{\sc iii}] 5007+
line ratio could be  twice solar. On the other hand, Fe is depleted 
from the gaseous phase by a factor $>$ 2, most probably  trapped 
into grains. Ratios of calculated to observed line ratios to \Hb
~indicate an average contribution of the broad line region to the 
observed \Hb ~of about 40\%.

\end{abstract}


\section{Introduction}

A large amount of information about the structure and
physical conditions of active galactic 
nuclei (AGN) has been obtained during the last decade
from the analysis of their optical emission-line spectrum 
(see the review of  \citealt{vcv00} 
and references therein). 
More recently, spectroscopic observations in the near- and 
mid-infrared have been added to that of the optical with enough 
resolution to reveal important aspects from both the narrow 
line and broad line regions (NLR and BLR, respectively; 
\citealt{sb01, kno01, rod02a, rod02b}) as well as 
from the circumnuclear environment \citep{tho00, lutz03, ver03}. Given that
the infrared region is substantially less affected by
extinction, it also allows one to probe depths unreachable with 
data at shorter wavelengths \citep{sb01, nag02}.

Among the different types of active galaxies, the narrow-line 
Seyfert 1 class (NLS1, \citealt{op85}) are special targets
because they share all optical properties of the Seyfert 1s except
the very narrow Balmer lines (FWHM$<$2000 km\,s$^{-1}$, by definition)
and the strong optical Fe\,{\sc ii} emission. In the X-rays region, NLS1 
are also peculiar because of the soft X-rays excess above the
hard X-rays power-law extrapolation that dominates below 1\,keV 
\citep{pun92, bbf96} and the very steep soft 
photon spectral index \citep{lei99b}.
 These extreme properties could derive from the fact of NLS1
being sources  with relatively small black hole masses 
(M$_{\rm BH}\sim$10$^{6-7}$~M$\odot$) and high Eddington ratios
L/L$_{edd} \sim$1-10 (e.g \citet{bor02, gru04}).

In the near-infrared (NIR), NLS1 usually present a rich emission-line 
spectrum \citep{rod02a, rod02b, gt02}. In addition, features associated to the BLR
such as the Fe\,{\sc ii} and the permitted H\,{\sc i} lines,
strongly blended in classical Seyfert 1 (Sy 1),
are easily separated in NLS1 because of the narrowness of their
BLR lines and the larger
separation in wavelength between the individual features. 
This, in turn, favors the  analysis of the forbidden line spectrum.
Weak features, usually lost amid strongly blended BLR lines in 
the optical region, are more  easily seen in the NIR spectrum. 
It is then interesting  to study the cases where the NLR lines are 
conspicuous in order to examine the physical conditions of the 
emitting gas and the  continuum properties to construct a coherent 
picture that explains the parameters driving the emission of the NLR 
in NLS1.

With the above in mind, a program to  consistently
study a selected sample of NLS1 galaxies, with emphasis
on the NIR information, was started with Ark 564 (\citealt{crv03}; hereafter Paper I). 
In this work, a similar analysis is 
applied to the NLS1 galaxy Mrk\,766(=NGC\,4253). 
Located at $z$ = 0.0129, it is classified as a barred SBa galaxy 
with optical overall orientation P.A=69$\degr$ \citep{mac90}. 
HST images of this object show filaments, wisps and irregular dust lanes 
around an unresolved nucleus \citep{mvt98}. 
Radio observations at 3.6\,cm, 6\,cm and 20\,cm 
\citep{uag95, nag99} show that the
radio source appears to be extended in both P.A. $\sim$27$\degr$
(on a scale of 0$\arcsec$.25) and P.A. 160$\degr$ (on
a scale of 0$\arcsec$3, \citealt{nag99}). In the optical, the
emission is extended \citep{gp96, mwt96} through a region of total size greater 
than that of the radio source. 

In X-rays, Mrk~766 is 
variable on a few hours time-scale and presents a strong 
soft X-ray excess \citep{mms93, lei96, mas03}. 
 \citet{bbf96} report one of the  flattest {\it ROSAT} 
soft X-ray photon spectral index for Mrk\,766 ($\alpha_X$ =-1.5)
in their sample of NLS1 galaxies, yet steeper than what most 
broad line Seyfert~1 or even some other NLS1 show.
\citet{gru01} give a $\alpha_X$ =-1.77 for this source, which is
exactly the mean value for the whole sample of soft X-ray
selected AGN reported in \citet{gru04}, not as flat as 
initially found. \citet{nan97} 
reported the detection of an Fe K$\alpha$ line in the X-ray 
spectrum. Moreover, recent {\it XMM-Newton} observations have caused
a debate over whether or not the soft X-ray spectrum arises
from a dusty warm absorber or 
relativistic emission lines \citep{mas03}.

In the ultraviolet, the spectrum of Mrk~766   looks also unusual
among Seyfert 1 galaxies  due to the weakness of both the ultraviolet
continuum and big blue bump \citep{wf93}. The deficit of ultraviolet photons
has been attributed to reddening \citep{pag99} and/or to absorption
\citep{wf93}. Actually, \citet{wf93} claim that there is no basic 
difference between the unabsorbed optical to X-ray continua emitted by 
Mrk~766 and by other Seyfert~1 galaxies.
The dusty scenario is 
supported by the optical polarization of Mrk\,766, of $\sim$2 per cent
increasing to the blue, due to scattering by
dust grains. The grains are probably located within the 
narrow line region given that the broad lines show more
polarization than the narrow lines \citep{good89}.

In the near-infrared, much of the information available for Mrk\,766
comes from imaging studies. \citet{alo98} report
that the surface brightness of this source at the 0.5--1~kpc radius
is 1-2\,mag\,arcsec$^{-2}$ brighter than that of normal galaxies.
Moreover, it shows one of the reddest {\it H-K} colours for the
0.5 and 1\,kpc diameter apertures of their sample. This result
is confirmed by the {\it J-} and {\it K'}-band imaging work of
\citet{mar99}, where it is shown that Mrk\,766
displays a double nuclear structure with a {\it (J-K')} color 
index redder than the rest of the galaxy. That structure seems to 
correspond to a feature delineated by the dust pattern that 
surrounds the innermost 2$\arcsec$ in the HST optical image 
\citep{mvt98}. Spectroscopically, Mrk\,766 have been poorly 
studied in this band. To our knowledge, the only observations of Mrk\,766 
published in the NIR region is the {\it K-}band spectroscopy of
\citet{gt02} and the {\it L-}band spectroscopy
of \citet{rv03}. In the former,
the spectrum clearly reveals Br$\gamma$ and probably the H$_{2}$ 
and [Si\,{\sc{vi}}] lines. In the later, the 3.3$\mu$m 
emission, attributed to polycyclic aromatic hydrocarbons 
(PAH), is detected.  The observation of PAH emission is usually
taken as unambiguous evidence of
starburst activity.   Interestingly,
 a strong starburst might be behind the high [O\,{\sc iii}]/\Hb\ 
ratio observed in Mrk~766, 
so  extreme \citep{gru04}
as to possibly exclude Mrk~766 from the NLS1 class, following the definition
 of NLS1s by \citet{op85}.

For all the above, Mrk\,766 arises as a very interesting
target to continue our analysis of the continuum and NLR
properties of NLS1 galaxies. To this purpose, NIR observations  
supplemented with archival UV - optical and mid-far infrared data will
be used in this work. This set of data is described in Sec.~\ref{obs}. The 
description and characterization of the observed NIR
continuum is described in Sec.~\ref{sednir}. A discussion about
the most important emission line features detected in the spectrum, both from the
BLR and NLR, is presented in Sec.~\ref{ler}. In a first 
approach  to apprehend the nature of Mrk~766, we compare its SED with that of NLS1 
galaxies (Sec.~\ref{SED}), and, particularly, with the SED of Ark~564, which
has been already modeled (Paper I ). This step will help at 
constraining the models for the evaluation of the emission-line 
ratios presented in Sec.~\ref{models}. After modeling the emission-line 
spectrum, the associated (and consistent) SED of the continuum 
is discussed in Sec.~\ref{modSED}. The concluding remarks appear in 
Sec.~\ref{fin}.

\section{Observations} \label{obs}

\begin{figure*}
\includegraphics[width=150mm]{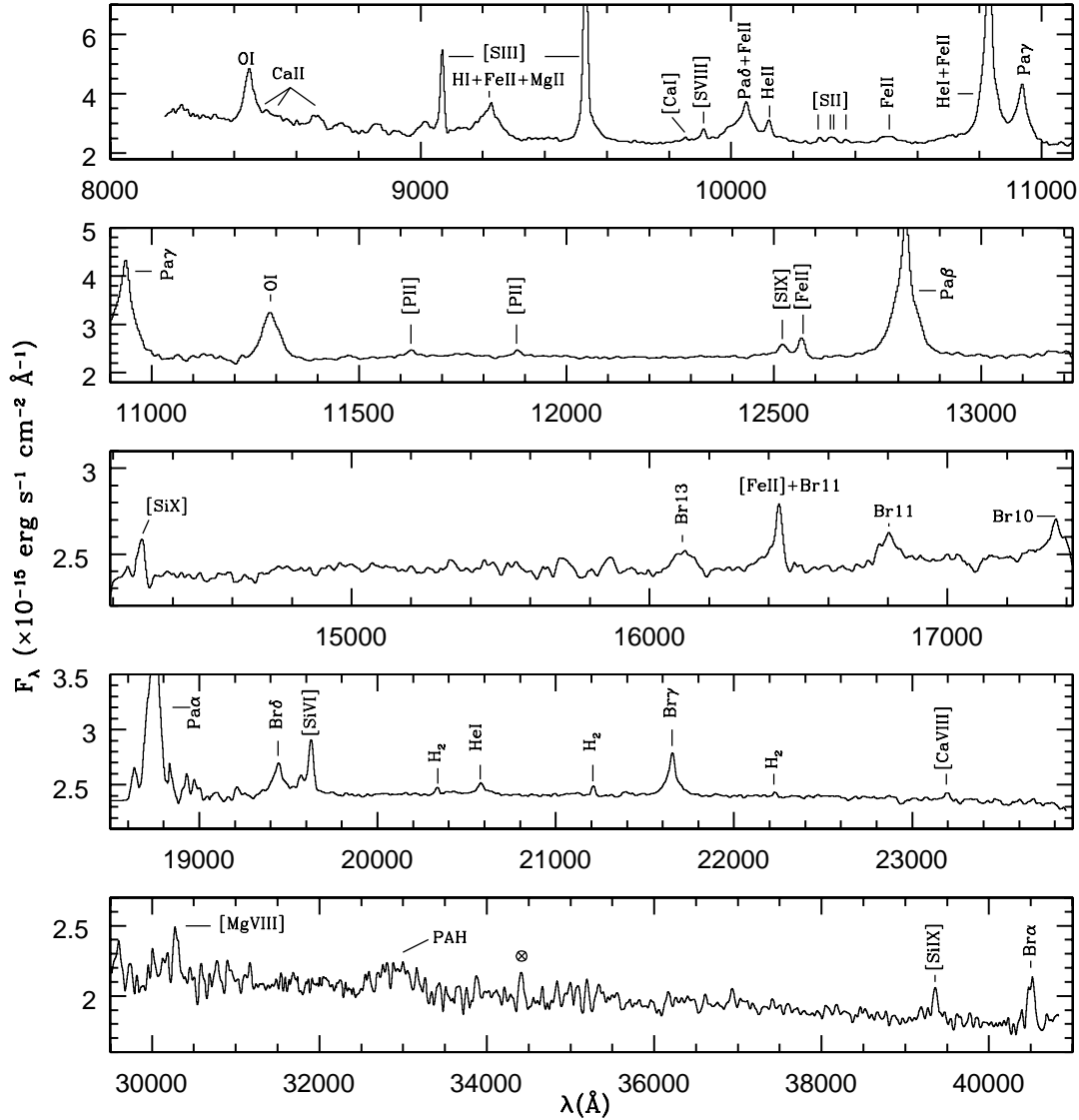}
\caption {Observed NIR spectrum of Mrk 766 in laboratory wavelengths. 
In order to easy the identification of the spectral features, the most 
important permitted and forbidden lines are marked.}\label{mrk766nir}
\end{figure*}

NIR spectra of Mrk~766 in the interval 0.8$-$2.4~$\mu$m were 
obtained at the NASA 3\,m Infrared Telescope Facility (IRTF) on 
the nights of April 21 and 25, 2002 (UT) using the
SpeX spectrometer \citep{ray03}. In addition, spectra covering
the interval 3-4~$\mu$m were also obtained with the same setup 
on the night of April 22 (UT). In both cases, the detector consisted
of a 1024$\times$1024 ALADDIN 3 InSb array with a spatial scale of
0.15$\arcsec$/pixel. Simultaneous wavelength coverage was obtained by means of
prism cross-dispersers.  A 0$\arcsec$.8$\times$15$\arcsec$ slit, 
oriented at the PA of 112$\degr$, was used during the observations. The resulting
spectral resolution was 360 \kms\ for both setups.
The observations were done nodding in an
ABBA source pattern with individual integration times of 120\,s and
30\,s for the short (0.8--2.4~$\mu$m) and long (3--4~$\mu$m) wavelength 
range, respectively.
The total on-source integration time was 1680\,s and 1080\,s for the April
21 and 25, respectively, and 900\,s for the 22 April observation. 
The A0\,V star HD79108  (V=6.13)  was observed 
after the target's to remove the telluric bands. It was also used to 
flux calibrate the spectra. The spectra were reduced following
standard procedures for NIR spectroscopy. They include subtracting
consecutive 2-D AB pairs to remove background, flat fielding
the resulting imaging, extracting the signal along the dispersion
axis, co-adding and calibrating.  The signal within the central 
0$\arcsec$.8$\times$1$\arcsec$ was summed up to obtain the nuclear 
spectrum. The 1$\arcsec$ extraction window along the slit, with 
centre at the peak continuum luminosity in K, is
in accord to the seeing of 0.$\arcsec$8 measured during the observations.
This window size corresponds to the inner 250~pc (H$_{\rm  o}$=75 km~s$^{-1}$~Mpc$^{-1}$)
of the AGN. No effort to extract information from the extended emission,
seen up to $\sim$400~pc NW and SE of the centre, was done.  
The reduction and extraction procedures were done using
the in-house software Spextool \citep{cvr04} provided by the IRTF
observatory. For telluric band correction
and flux calibration, a script also provided by IRTF, designed to this
purpose \citep{vcr03} was employed. Afterward, the two spectra in the short 
wavelength range were averaged to form a single 0.8$-$2.4 spectrum.

Figure~\ref{mrk766nir} show the final  rest-frame
NIR spectrum of Mrk~766.  Note that within the intervals
1.74~$\mu$m--1.86~$\mu$m and 2.4~$\mu$m--2.9~$\mu$m, the atmospheric
transmission drops to zero. For this reason, these two regions are
missing in Figure~\ref{mrk766nir}.  As can be seen, the spectrum is dominated
by permitted emission lines of H\,{\sc i}, He\,{\sc i},
O\,{\sc i} and Fe\,{\sc ii}. Also prominent are the forbidden lines of
[S\,{\sc iii}] $\lambda\lambda$9068, 9531, [Si\,{\sc vi}] 1.963~$\mu$m, [Mg\,{\sc viii}]
3.03~$\mu$m and [Si\,{\sc ix}] 3.936~$\mu$m. Note also
the bump centred at 3.3~$\mu$m, attributed to emission of PAH
molecules \citep{rv03}. It represents an unambiguous signature of
starburst activity within the central 250~pc of this NLS1 galaxy.

Fluxes of the identified NIR lines measured from
our spectrum are listed in Table~\ref{tab1}. They were calculated
by fitting a Gaussian to the observed profile and then integrating 
the flux under the curve. The LINER software \citep{po93}  
was used for this purpose. 

 Archival {\it HST} data taken with the Space 
Telescope Imaging Spectrograph (STIS) on 11 April 2000 (UT) were also 
used in order to measure the flux of the
most important UV lines of Mrk\,766. The spectra were obtained through the
52$\arcsec \times 0.2\arcsec$ slit and the low-resolution (R$\sim$750)
G140L and G230L gratings. The spectra cover the rest-frame wavelength range 
1100--1690\AA\ and  1550--3200\AA, respectively. The final rectified
wavelength calibrated x2d spectrum produced by the {\it HST} pipeline
was used to extract the 1-D spectrum. The extraction window in the
spatial direction was the same used in the NIR data. The IRAF package
was used to this purpose.

In the optical region, fluxes for the most important lines 
were taken from \citet{gp96} (hereafter, GP96). They report 
relatively strong Fe\,{\sc ii} emission and coronal lines, as well as
extended optical emission through a region the total size of 
8" (about 2 kpc) along the stellar bar. Because of the high quality of
the data and resolution, that matches ours', the nuclear fluxes of 
GP96 will be used through this work. 

Finally, the fluxes of the emission lines from UV to NIR were
derredened for Galactic and internal extinction. The value of E(B-V) 
corresponding to our Galaxy is 0.02 \citep{sfd98}. The internal
extinction E(B-V)=0.5$\pm$0.1,  applied to the data, was determined 
from the average of the
extinction values derived  from the flux ratio of the   narrow 
components of Pa$\beta$/Pa$\gamma$, E(B-V)=0.69, and that determined
by GP96 [E(B-V)=0.38].
 The value found from the Pashen lines agrees with that derived 
from the Balmer decrement \citep{gru98}, indicating that Mrk~766 is indeed
highly reddened in the optical/UV. Note that higher values of extinction were
also found, for example, from the flux ratios [Fe\,{\sc ii}] 1.257$\mu$m/1.644$\mu$m 
[E(B-V)=1.2] and Br$\gamma$/Pa$\beta$ [E(B-V)=1.15]. However,
we rejected the latter two values because it is not certain if 
the bulk of the NLR gas is affected by such
large values of extinction, mainly if we take into account that 
the optical spectrum of Mrk\,766 displays prominent [Ne\,{\sc v}]  3346, 3426~\AA 
lines (see Figure 3 of GP96), which are located in the blue end of 
the spectrum. Table~\ref{derfluxes} list the
dereddened optical and NIR narrow line fluxes. Note that the
extinction law of \citet{ccm89} was applied in that
correction. 

 The choice of the \citet{ccm89} extinction law
is due to the fact that we are assuming that the
dust affecting the NLR is located in a non-nuclear environment. It is yet possible that
the observed continuum in the UV-optical region is affected by the 
same amount of extinction, but if there is an additional nuclear dust
component, it is likely that the \citet{ccm89} law
is no longer valid. The main support to this is the lack of the $\lambda$2175 dip
in the UV continuum shape. This is in accord to recent claims by
\citet{gas04}, who provided 
substantial evidence of a relative lack of small grains in the nuclear 
dust, which would explain the absence of the $\lambda$2175 dip in most
AGN spectra. They proposed a reddening curve fairly universal, being
quite flat in the UV but selective in the optical. Due to the lack
of enough continuum points in the optical region that adequately
connects the NIR with the {\it HST} data, we feel that any correction 
to the nuclear UV continuum is highly uncertain. Since our interest is 
concentrated in the NLR gas and the reprocessed continuum by this
gas, we feel that this correction does not affect our main
results and conclusions. 

\begin{table*}
\begin{center}
\caption{Observed NIR emission line fluxes measured in Mrk\,766} \label{tab1}
\begin{tabular}{lccc}
\hline \hline
    & Flux  &   & Flux \\
Line  & x10$^{-15}$ erg cm$^{-2}$ s$^{-1}$  & Line  & x10$^{-15}$ erg cm$^{-2}$
s$^{-1}$ \\
\hline 
Br$\alpha$ 4.052$\mu$m & 31.3$\pm$4.70 &  He{\sc ii} (b$^{*}$) ~1.162$\mu$m & 2.94$\pm$0.81 \\
\lbrack Si{\sc ix}\rbrack~3.935$\mu$m & 17.06$\pm$3.60 & \lbrack P{\sc ii}\rbrack
~1.146$\mu$m & 1.44$\pm$0.29 \\
\lbrack Mg{\sc viii}\rbrack~3.027$\mu$m & 26.6$\pm$8.06 & O{\sc i} ~1.128$\mu$m  &
48.86$\pm$2.40 \\
\lbrack Ca{\sc viii}\rbrack~2.321$\mu$m & 2.47$\pm$0.60 & Pa$\gamma$ (n$^{*}$)
~1.937$\mu$m & 13.05$\pm$0.77 \\ 
H$_{2}$ ~2.223$\mu$m       & 0.81$\pm$0.15 & Pa$\gamma$ (b$^{*}$) ~1.937$\mu$m &
87.30$\pm$3.07  \\
Br$\gamma$ (n$^{*}$) 2.157$\mu$m & 7.40$\pm$0.35 & Fe{\sc ii} ~1.086$\mu$m     &
50.6$\pm$2.40  \\
Br$\gamma$ (b$^{*}$) 2.157$\mu$m & 20.0$\pm$2.26 & He{\sc i} ~1.083$\mu$m &
216.4$\pm$1.53 \\
H$_{2}$ ~2.121$\mu$m       & 2.36$\pm$0.24 & Fe{\sc ii} ~1.050$\mu$m  &
14.86$\pm$1.96 \\
He{\sc i} (n$^{*}$) ~2.058$\mu$m   & 2.03$\pm$0.20 & \lbrack S{\sc ii}\rbrack ~1.037$\mu$m
& 1.67$\pm$0.46 \\
He{\sc i} (b$^{*}$) ~2.058$\mu$m   & 6.09$\pm$1.10 & \lbrack S{\sc ii}\rbrack ~1.033$\mu$m
& 2.11$\pm$0.46 \\
H$_{2}$ ~2.033$\mu$m        & 1.73$\pm$0.22  & \lbrack S{\sc ii}\rbrack ~1.032$\mu$m
& 3.38$\pm$0.46 \\
\lbrack Si{\sc vi}\rbrack ~1.963$\mu$m & 15.6$\pm$0.94 & \lbrack S{\sc ii}\rbrack
~1.028$\mu$m & 2.91$\pm$0.46 \\ 
H$_{2}$ ~1.957$\mu$m        & 4.39$\pm$0.39 & Fe{\sc ii} ~1.017$\mu$m      &
6.23$\pm$2.10 \\
\lbrack Fe{\sc ii}\rbrack ~1.634$\mu$m & 8.20$\pm$0.40 & He{\sc ii} (n$^{*}$) ~1.012$\mu$m
 & 18.57$\pm$2.02 \\
\lbrack Si{\sc x}\rbrack ~1.430$\mu$m & 6.47$\pm$0.54 & He{\sc ii} (b$^{*}$) ~1.012$\mu$m 
& 5.68$\pm$0.70 \\
Pa$\beta$ (n$^{*}$) ~1.281$\mu$m & 27.15$\pm$0.50 & Pa$\delta$ (n$^{*}$) ~1.004$\mu$m  &
4.82$\pm$0.69  \\
Pa$\beta$ (b$^{*}$) ~1.281$\mu$m & 117.8$\pm$1.80 & Pa$\delta$ (b$^{*}$) ~1.004$\mu$m  &
68.7$\pm$2.7 \\
\lbrack Fe{\sc ii}\rbrack ~1.257$\mu$m & 7.69$\pm$0.70 & Fe{\sc ii} ~0.999$\mu$m    
 & 21.98$\pm$2.13 \\
\lbrack S{\sc ix}\rbrack ~1.252$\mu$m & 9.42$\pm$0.96 & \lbrack S{\sc viii}\rbrack
~0.991$\mu$m & 10.0$\pm$1.00 \\
\lbrack P{\sc ii}\rbrack ~1.188$\mu$m & 2.19$\pm$0.29 & \lbrack Ca{\sc i}\rbrack
~0.985$\mu$m & 5.90$\pm$1.34 \\
He{\sc ii} (n$^{*}$) ~1.162$\mu$m & 1.40$\pm$0.24 & \lbrack S{\sc iii}\rbrack ~0.953$\mu$m
& 110.0$\pm$0.90 \\  
\hline
\end{tabular}
\end{center}
{\it n} and {\it b} stand for the narrow and broad components, respectively,
of the permitted emission lines.  
\end{table*}

\begin{table*}
\centering
\caption{Line fluxes in Mrk\,766 corrected for E(B-V)=0.5 and normalized to F$_{5007+}\times 100$
\label{derfluxes}}
\footnotesize{
\begin{tabular}{lcclcc}
\hline \hline
    & Flux  & F$_{\lambda}$/F$_{5007}$ & &  Flux & F$_{\lambda}$/F$_{5007}$   \\
Line  & 10$^{-14}$ erg cm$^{-2}$ s$^{-1}$ &  &   Line & 10$^{-14}$ erg cm$^{-2}$
s$^{-1}$ &     \\
\hline
\lbrack Si{\sc ix}\rbrack~3.935$\mu$m   & 1.87$\pm$0.39 &    1.10$\pm$0.25  &
\lbrack S{\sc ii}\rbrack~6717 \AA & 4.14$\pm$0.33 &    2.45$\pm$0.26  \\
\lbrack Mg{\sc viii}\rbrack~3.027$\mu$m & 2.94$\pm$0.88 &    1.74$\pm$0.54  & He{\sc
i}~6678 \AA                      & 2.00$\pm$0.22 &    1.18$\pm$0.16  \\
\lbrack Ca{\sc viii}\rbrack~2.321$\mu$m & 0.30$\pm$0.07 &    0.17$\pm$0.04  &
\lbrack N{\sc ii}\rbrack~6584 \AA       & 18.01$\pm$1.08&    10.66$\pm$0.99 \\
He{\sc i}     ~2.058$\mu$m              & 0.96$\pm$0.13 &    0.57$\pm$0.09  &
H$\alpha$ ~6563 \AA                     & 281.1$\pm$16.9&    166.3$\pm$15.5 \\
\lbrack Si{\sc vi}\rbrack ~1.963$\mu$m  & 1.87$\pm$0.11 &    1.11$\pm$0.10  &
\lbrack N{\sc ii}\rbrack~6548 \AA       & 4.00$\pm$0.12 &    2.37$\pm$0.18  \\
\lbrack Fe{\sc ii}\rbrack ~1.634$\mu$m  & 1.05$\pm$0.05 &    0.62$\pm$0.05  & Fe{\sc
x}\rbrack~6375 \AA               & 0.45:         &    0.27$\pm$0.02  \\
\lbrack Si{\sc x}\rbrack ~1.430$\mu$m   & 0.91$\pm$0.07 &    0.54$\pm$0.06  &
\lbrack O{\sc i}\rbrack~6363 \AA           & 1.28$\pm$0.12 &    0.76$\pm$0.09  \\
Pa$\beta$~1.281$\mu$m                   & 21.34$\pm$0.21&    12.63$\pm$0.91 &
\lbrack O{\sc i}\rbrack~6300 \AA        & 2.80$\pm$0.22 &    1.66$\pm$0.18  \\
\lbrack Fe{\sc ii}\rbrack ~1.257$\mu$m  & 1.12$\pm$0.10 &    0.66$\pm$0.08  &
\lbrack Fe{\sc vii}\rbrack~6086 \AA     & 2.40$\pm$0.12 &    1.42$\pm$0.12  \\
\lbrack S{\sc ix}\rbrack ~1.252$\mu$m   & 1.33$\pm$0.13 &    0.79$\pm$0.10  & He{\sc
i} ~5876 \AA                     & 13.74$\pm$1.10&    8.13$\pm$0.87  \\
\lbrack P{\sc ii}\rbrack ~1.188$\mu$m   & 0.35$\pm$0.05 &    0.21$\pm$0.03  &
\lbrack N{\sc ii}\rbrack~5755 \AA       & 0.37$\pm$0.06 &    0.22$\pm$0.04  \\
He{\sc ii}     ~1.162$\mu$m             & 0.68$\pm$0.13 &    0.40$\pm$0.08  &
\lbrack Fe{\sc vii}\rbrack~5721 \AA     & 1.33$\pm$0.12 &    0.79$\pm$0.09  \\
\lbrack P{\sc ii}\rbrack ~1.146$\mu$m   & 0.23$\pm$0.05 &    0.13$\pm$0.03  &
\lbrack O{\sc i}\rbrack~5577 \AA        & 0.33$\pm$0.07 &    0.20$\pm$0.04  \\
Pa$\gamma$ ~1.937$\mu$m                 & 12.27$\pm$0.37&    7.26$\pm$0.56  &
\lbrack O{\sc iii}\rbrack~5007 \AA      & 169.3$\pm$11.9&    100.0$\pm$7.10 \\
He{\sc i} ~1.083$\mu$m                  & 36.02$\pm$0.36&    21.31$\pm$1.53 &
H$\beta$~4861~\AA                               & 82.93$\pm$2.95&    49.$\pm$4.  \\
\lbrack S{\sc ii}\rbrack ~1.037$\mu$m   & 0.29$\pm$0.08 &    0.17$\pm$0.05  &
\lbrack Ar{\sc vi}\rbrack~4740 \AA      & 0.67$\pm$0.09 &    0.39$\pm$0.06  \\
\lbrack S{\sc ii}\rbrack ~1.033$\mu$m   & 0.36$\pm$0.08 &    0.21$\pm$0.05  &
\lbrack Ar{\sc vi}\rbrack~4711 \AA      & 0.63$\pm$0.09 &    0.37$\pm$0.06  \\
\lbrack S{\sc ii}\rbrack ~1.032$\mu$m   & 0.57$\pm$0.08 &    0.34$\pm$0.05  & He{\sc
ii}~4686 \AA                     & 7.34$\pm$0.51 &    4.34$\pm$0.43  \\
\lbrack S{\sc ii}\rbrack ~1.028$\mu$m   & 0.53$\pm$0.09 &    0.32$\pm$0.06  &
\lbrack O{\sc iii}\rbrack~4363 \AA      & 3.47$\pm$0.59 &    2.05$\pm$0.38  \\
He{\sc ii} ~1.012$\mu$m                 & 4.27$\pm$0.38 &    2.53$\pm$0.29  &
H$\gamma$ 4341 \AA                      & 34.82$\pm$3.13&    20.6$\pm$2.36  \\
Pa$\delta$  ~1.004$\mu$m                & 12.94$\pm$0.52&    7.66$\pm$0.62  &
\lbrack S{\sc ii}\rbrack~4070 \AA       & 2.54$\pm$0.46 &    1.50$\pm$0.29  \\
\lbrack S{\sc viii}\rbrack ~0.991$\mu$m & 1.87$\pm$0.19 &    1.11$\pm$0.14  &
\lbrack Ne{\sc iii}\rbrack+H{\sc i}~3968 \AA & 6.27$\pm$0.63&  3.71$\pm$0.46\\
\lbrack Ca{\sc i}\rbrack ~0.985$\mu$m   & 1.04$\pm$0.24 &    0.62$\pm$0.15  &
H8+He{\sc i}~3888 \AA                   & 3.60$\pm$0.61 &    2.13$\pm$0.39  \\
\lbrack S{\sc iii}\rbrack ~0.953$\mu$m  & 20.54$\pm$0.21&    12.16$\pm$0.87 &
\lbrack Ne{\sc iii}\rbrack~3868 \AA     & 20.41$\pm$1.63&    12.08$\pm$1.29 \\
\lbrack Ni{\sc ii}\rbrack~7378 \AA      & 0.40$\pm$0.04 &    0.24$\pm$0.03  &
H9+He{\sc ii}~3835 \AA                  & 1.09$\pm$0.51 &    0.65$\pm$0.31  \\
\lbrack O{\sc ii}\rbrack~7320+7330 \AA  & 1.87$\pm$0.11 &    1.11$\pm$0.10  &
\lbrack Fe{\sc vii}\rbrack~3758 \AA     & 4.00$\pm$0.92 &    2.37$\pm$ 0.57 \\
\lbrack Ar{\sc iii}\rbrack~7135 \AA             & 3.20$\pm$0.22 &    1.89$\pm$0.19 
& \lbrack O{\sc ii}\rbrack~3727 \AA       & 22.28$\pm$1.56&    13.18$\pm$1.31
\\
He{\sc i}~7065 \AA                      & 3.07$\pm$0.21 &    1.82$\pm$0.18  &
\lbrack Ne{\sc v}\rbrack~3426 \AA       & 28.01$\pm$1.68&    16.58$\pm$1.54 \\
\lbrack Ar{\sc vi}\rbrack~7006 \AA              & 1.60$\pm$0.11 &    0.95$\pm$0.09 
& \lbrack Ne{\sc v}\rbrack~3345 \AA       & 10.0$\pm$0.80 &    5.92$\pm$0.63
\\
\lbrack S{\sc ii}\rbrack~6731 \AA               & 5.34$\pm$0.32 &    3.16$\pm$0.29 
& ... & ... \\
\hline
\end{tabular}}
\end{table*}

\section{The NIR Continuum} \label{sednir}

\subsection{The observed continuum}

\begin{figure}
\includegraphics[width=88mm]{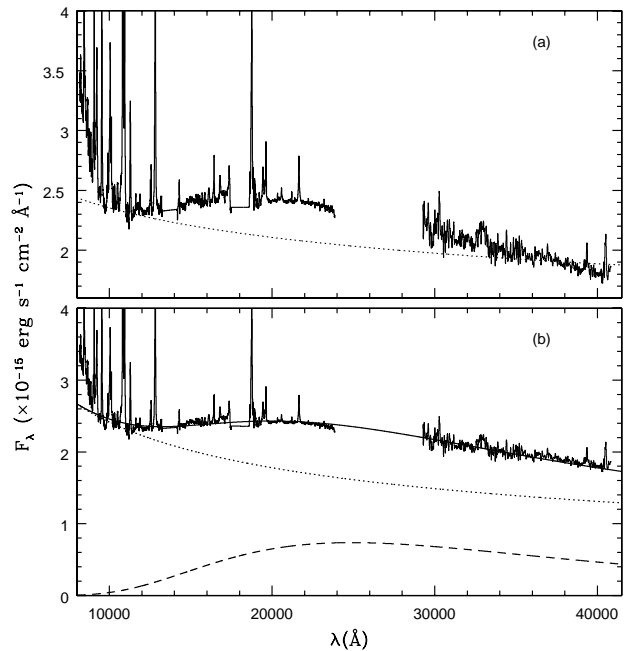}
\caption {(a) NIR SED observed in Mrk\,766. The dotted line is best-fitted
power-law function. Note the strong excess of emission over the power-law,
in the interval 12000--32000 \AA. Note also that the power-law
overpredicts the continuum emission lonward of 38000 \AA. (b)
NIR SED modeled with a composite function: a power-law (dotted lines),
of spectral index $\alpha$=-0.44, as found by \citet{al03}, plus a blackbody
distribution (dashed line) of $T_{\rm bb}$=1200~K. The thick line
represent the sum of both components. \label{powl}}

\end{figure}

The continuum emission displayed by Mrk\,766 in the interval 
0.8$-$4~$\mu$m is complex (see Figure~\ref{powl}). At least 
three main features can be 
easily identified in its spectrum: (i) a 
steep continuum blueward of 1~$\mu$m, very similar to the the 
so-called small blue bump in the optical region observed in 
AGN around 4000 \AA; (ii) a turnover of the continuum at $\sim$1.1~$\mu$m. 
The continuum emission raises longward up to $\sim$2.2~$\mu$m, 
where it reaches a maximum; (iii) a steep decline of the 
continuum flux with wavelength, longward of 2.2~$\mu$m. 

Few AGN at a similar wavelength coverage and
spectral resolution in the NIR region are available in the 
literature, so a meaningful comparison of the continuum 
observed in Mrk\,766 with that of other AGN is not possible. 
However, data published for the NLS1s galaxies I\,Zw\,1
\citep{rudy00}, Mrk\,478 \citep{rud01} and Ark\,564, Ton\,S\,180, 1H\,1934-063
and Mrk\,1044 \citep{rod02a} show that
the turnover of the continuum near 1.1$\mu$m is
a common feature and may represent the shift
from the non-thermal continuum emitted by the active nucleus 
to the stellar and thermal dust emission from
the circumnuclear region that dominates at longer wavelengths. 
The rapid rise of the continuum in the J and H bands with 
a peak at the K-band, and then its slow decline, has not been
reported before, to our knowledge, for an AGN but I\,Zw\,1. 
Note, however, that the I\,Zw\,1 spectrum of \citet{rudy00} ends 
at 2.40\,$\mu$m. 

In order to characterize the NIR continuum emission in
Mrk\,766, a power-law of the form 
F$_{\lambda} \propto \lambda^{\alpha}$ was fitted to the
data. The task {\it nfit1d} of the STSDAS package of IRAF
was used to this purpose. Special care was taken to not include
emission lines in the spectral windows used
in the fit. The result is shown in the upper panel of
Figure~\ref{powl}. The derived spectral index $\alpha$ was -0.16, 
significantly flatter than the average value of -0.64 
found for the AGN sample of \citet{em86} and also
flatter than the value of -0.44 found by \citet{al03}
in the interval 1$-$16$\mu$m for Mrk\,766.
It can be seen from Fig.~\ref{powl} that the 
fitted function overpredicts the continuum emission redward of 
4$\mu$m. In addition, it shows that a single power-law 
cannot describe the continuum emission in the 1$-$4$\mu$m 
region. There is an excess of emission over the
power-law that in  no way follows a simple 
function.

A closer look to the overall shape of the excess of emission 
over the power-law, in the wavelength range 1--4~$\mu$m, 
suggests that it approaches to that of a blackbody distribution. In order
to test this hypothesis, we fitted a composite function 
(power-law plus blackbody) to the observed NIR SED. 
In the fit, the spectral index of the power-law was constrained to the value
$\alpha$=-0.44, found by \citet{al03} while the temperature and
amplitude of the blackbody function were left as free parameters.
The result, displayed in the lower panel of Figure~\ref{powl}, show that
the composite function, with a blackbody of 
temperature $T_{\rm bb}$=1200~K,  adequately describes the
continuum distribution. Integrating the flux over these two
components in the observed interval gives  F$_{\rm bb}$=1.55$\times$10$^{-11}$
erg\,cm$^{-2}$\,s$^{-1}$ and F$_{\rm PL}$=5.60$\times$10$^{-11}$
erg\,cm$^{-2}$\,s$^{-1}$  for the blackbody and power-law
functions, respectively. It means that in the NIR region, the
former component accounts for  28\% of the total continuum flux.

What is the origin of each of the two continuum components, 
both emitted in the inner 250\,pc of the AGN? Given the composite
nature of Mrk\,766, harboring both a dusty active nucleus and
a circumnuclear starburst, there are several possibilities. The
power-law continuum can be associated to continuum emission from
the AGN plus some contribution from circunnuclear stellar population
and dust. Note that this power-law cannot simply be regarded as
an extrapolation to the power-law used to describe the
continuum in the UV and optical region. This is because in the
optical, the continuum displayed by Mrk\,766 is essentially
flat, even in the UV region (see Figure~3 of GD96). This
suggest that the continuum emission, at shorter wavelengths,
is strongly reddened. Under the assumption that the power-law 
found in the NIR carries a significant contribution from the AGN,
it can be used to find out an upper limit to the extinction
affecting the UV and optical region. We determined that 
an at least a E(B-V)=0.3 to the continuum is necessary in
order to a power-law of spectral index $\alpha$=-0.44
can adequately represents the AGN continuum in the optical
region. 

Regarding the nature of the excess of emission well described
by a black body function, it could be related to a compact
dust/molecular thick torus \citep{gd94} or result from 
emission from hot dust (T$>$1000 K). This reasoning follows
similar distributions found in other AGN by \citet{ma98} 
and \citet{ma00}.

\subsection{The hot dust hypothesis}

The excess of emission over the power-law distribution
in the NIR region and its adequate representation obtained 
after adding a blackbody distribution suggest that it can
be due to hot dust grains. The temperature derived
from the fit ($T_{\rm bb}$=1200~K) is near to the peak value for the 
hottest possible grains, i.e., graphite grains at their 
evaporation temperature, T$\sim$1500\,K, and higher
than the sublimation temperature of  silicate
grains (T$\sim$1000~K; \citealt{gd94}). 
Considering that our spatial resolution is limited to 
$\sim$250 pc, it is very likely that dust at higher temperatures 
exists closer to the central source, ruling out the possibility
of silicates as the main component of the nuclear dust grains.
This result does not exclude that silicate grains can 
exist within the NLR of Mrk~766. They still can be present
within the observed region but a much lower temperature. 
Recalling that HST observations of Mrk\,766 reveals 
filaments, wisps and irregular dust lanes around the 
unresolved nucleus \citep{mvt98} and that the 
optical polarization of Mrk\,766 is due to scattering from
dust grains \citep{good89}, the dusty NLR scenario
for Mrk\,766 is strongly supported by the NIR data.

Using the temperature of the blackbody distribution derived
above as the average temperature of the graphite grains and 
a K band flux of 1.03$\times$10$^{-25}$ erg s$^{-1}$ 
cm$^{-2}$ Hz$^{-1}$ at 2.2\,$\mu$m found for the blackbody 
component after subtracting the power-law,
we can roughly estimate the dust mass associated with the 
excess of emission. Following \citet{bar87}, the infrared 
spectral luminosity, in ergs s$^{-1}$ Hz$^{-1}$, of an 
individual graphite grain is:

\begin{equation}
L_{\nu,\mathrm{ir}}^{\mathrm{gr}}=4\pi~a^{2}~\pi Q_{\nu} B_{\nu}(T_{\mathrm{gr}})
\end{equation}

where $a$ is the grain radius, $Q_{\nu}= q_{\mathrm{ir}} \nu^{\gamma}$ is
the absorption efficiency of the grains and $B_{\nu}(T_{\mathrm{gr}})$ is the
Planck function for a grain of temperature $T_{\mathrm{gr}}$. Adopting,
as in \citet{bar87}, a value of $a$=0.05$\mu$m for graphite grains
and $Q_{\nu}$=0.058 and setting $T_{\mathrm{gr}}$=1200~K, we find
$L_{\nu,\mathrm{ir}}^{\mathrm{gr}}$=9.29$\times$10$^{-18}$ 
ergs s$^{-1}$ Hz$^{-1}$.  

The total number of emitting grains (hot dust) can be approximated
as:

\begin{equation}
N_{\mathrm HD} \approx \frac{L_{\mathrm NIR}}{L_{\nu,\mathrm{ir}}^{\mathrm{gr}}}.
\end{equation}

Finally, for graphite grains, with density $\rho_{\rm g}$=2.26 g~cm$^{-3}$,

\begin{equation}
M_{\mathrm HD} \approx \frac{4\pi}{3} a^{3} N_{\rm HD} \rho _{\mathrm g}
\end{equation}

Taking Mrk~766 at a distance of 51.6 Mpc  (z=0.0129, H$_o$=75 \kms/Mpc), we obtained 
$N_{\mathrm HD}$=3.54$\times 10^{45}$ and 
$M_{\mathrm HD}$=2.1$\times 10^{-3}$~M$\odot$. 

The mass of hot dust derived for Mrk~766 is similar to
that found in NGC~3783 (2.5$\times 10^{-3}$~M$\odot$, 
\citealt{gla92}) and NGC~1068 (1.1$\times 10^{-3}$~M$\odot$, 
\citealt{ma00}); larger than that derived for NGC~1566 
(7$\times 10^{-4}$~M$\odot$, \citealt{bar92}) and
NGC~4593 (5$\times 10^{-4}$~M$\odot$, \citealt{san95})
and significantly smaller than that of NGC~7469 
(0.05~M$\odot$, \citealt{ma98}) and Fairall~9 (0.09~M$\odot$, 
\citealt{cwg89}). Note that for all the above objects
but Mrk~766, the mass of hot dust was derived from
photometric  observations. The spectral resolution of our data
offers the possibility of examining a spectral feature 
not previously observed in such a detail. 

Regarding to the location of the hot dust, our spectral
resolution imply that it must reside within the inner 250\,pc from
the centre,  since this is the region size covered by 
the nuclear NIR spectrum employed in this work (see \S ~\ref{obs}).
The high temperature found for the dust, however, 
allows us to impose a tighter constraint to the location of the
emitting region: it would be located in the central 100\,pc. This value
is deduced by inspecting Fig.~3 of \citet{ma98}, where a
plot of dust temperature as function of distance from
the central source was constructed for NGC~7469. 
Note that in this reasoning, we have assumed that the dust in Mrk\,766
has similar properties to that of NGC~7469. Further support
to this value can be obtained following the results 
for NGC~1068 \citep{ma00}. These authors concluded that
hot dust ($T_{\mathrm{gr}}$=1500~K) should be extremely
confined and located at a radius less than 4~pc. 
If that is also the case for Mrk~766, it is most likely
that the excess of emission in the NIR is associated
to the torus than to circumnuclear dust. Only high
resolution spectroscopy will be able to disentangle
this possibility. Whatever the case, our results 
provides the first direct spectroscopic evidence of hot
dust in AGN and show the potential that NIR spectroscopy
has at unveiling that component.

\section{The line emitting regions} \label{ler}

\begin{figure*}
\includegraphics[width=120mm]{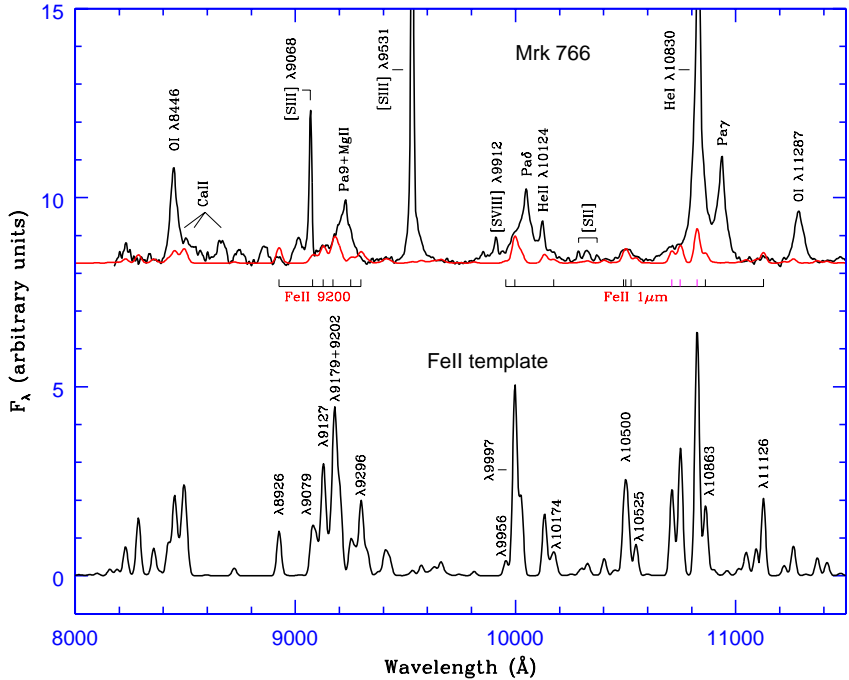}
\caption {Top: Mrk\,766 spectrum with the most important 
emission lines identified in the region between 0.8$-$1.1~$\mu$m.
The continuum emission has been subtracted and the resultant
spectrum displaced by an arbitrary amount for displaying purposes.
 The thick line is the best empirical
Fe\,{\sc ii} template derived from the models of \citet{sp03}.
Note the good agreement between the template and most of
Fe\,{\sc ii} features, mainly the 9200\AA\ blend and the 1~$\mu$m 
lines. Bottom: Fe\,{\sc ii} Empirical template with the most
important lines identified. Nothe that in order to ease the 
identification of individual lines, we show the template
before being convolved with
a Gaussian that matches the width of individual permitted lines
of Mrk~766. \label{feiilines}}
\end{figure*}

A careful inspection to the spectrum of Mrk\,766 shown in
Fig.~\ref{mrk766nir} evidences 
prominent permitted emission lines of Fe\,{\sc ii},
H\,{\sc i}, He\,{\sc i} and O\,{\sc i} and bright
forbidden lines of [S\,{\sc iii}]. These features carry
most of the energy radiated in the form of emission lines
in the 0.8$-$1.3$\mu$m interval. In addition, several coronal 
lines with ionization potential (IP) of up to 350 eV,
some of them previously observed in just a handful
of AGN, are detected. Such is the case of
[S\,{\sc viii}] 0.991~$\mu$m (IP=280.9 eV), [S\,{\sc ix}]
1.252~$\mu$m (IP=328.3 eV), [Si\,{\sc vi}] 1.963~$\mu$m
(IP=166.7 eV), [Si\,{\sc ix}] 3.936~$\mu$m (IP=303.17 eV),
[Si\,{\sc x}] 1.143~$\mu$m (IP=351.1 eV) and [Mg\,{\sc viii}]
3.03~$\mu$m (IP=224.95). Low ionization lines, with IP$<$13.6
eV, are also present (i.e., [C\,{\sc i}], [Fe\,{\sc ii}]
and [S\,{\sc ii}]).

The simultaneous observation of very high
and neutral/low ionization lines imply that a wide variety of
physical conditions must coexist within the
central few hundred parsecs of this object. In the following 
sections, with the help of ionization models, we will
extract information about the structure and physical
state of the NLR, taking as reference, the observed continuum and
emission line spectrum. A detailed analysis 
of the physical conditions and modeling of the BLR
features will be discussed in a separate paper 
(Rodr\'{\i}guez-Ardila et al. 2004, in preparation).
Here, we will briefly mention the most important
features identified in the BLR spectrum of Mrk~766.

Few AGN with positive detection of NIR Fe\,{\sc ii} lines 
have been reported in the literature so far 
\citep{rudy00, rud01, nag02, rod02a}. Here, we report the
identification of the Fe\,{\sc ii} 9200~\AA\ blend and
the Fe\,{\sc ii} 1\,$\mu$m lines\footnote{It is known under this name 
the set of Fe\,{\sc ii} lines located 
between 1$\mu$m and 1.2$\mu$m, among which 0.9997$\mu$m, 1.0171$\mu$m, 
1.050$\mu$m, 1.0863$\mu$m and 1.1126$\mu$m are the strongest ones} 
in Mrk~766,  as is shown in Fig.~\ref{feiilines}. In the upper part
of the Figure, we have marked the strongest transitions identified.
Overplotted, is the empirical template we derived. This work was done 
with the help of the NIR Fe\,{\sc ii} line list and models
of \citet{sp03}. A zoom of the empirical template, plotted 
at the bottom, was indeed constructed using the position and 
relative intensity of the lines listed by \citet{sp03}. We 
then adjusted by eye most of the line intensities in disagreement
with the model in order to get a better approximation to the 
observed spectrum.    

The detection of  
NIR Fe\,{\sc ii} lines provides important information on the
mechanism driving the formation of this ion. Primary cascade 
lines descending from the upper 5$p$
levels to the lower $e^{4}D$ and $e^{6}D$ terms, thought to 
be pumped by Ly$\alpha$ fluorescence \citep{jj84, sp98, sp03}, 
are located in the region between 8500\,\AA\ and 9500\,\AA. 
Their detection, as in Mrk~766, supports this mechanism as a 
competing one in the formation of the Fe\,{\sc ii}. Moreover, the  
1$\mu$m lines can also be pumped by Ly$\alpha$ fluorescence
(see Figure~16 of \citealt{sp03}).
The energy levels that are populated after
their emission originate nearly 50\% of the optical
Fe\,{\sc ii}. This de-excitation channel has been
overlooked for years and should be included in
modeling the Fe\,{\sc ii} spectrum. Given the importance 
that the Fe\,{\sc ii}  lines have in AGN studies, a paper 
describing the detection of NIR transitions and the
physical information that can be derived for a sample of AGN,
including Mrk~766, is currently in preparation.

In regard to the NLR, the width of the lines provide important 
clues about the velocity field of the emitting gas. This 
information will be useful for the models that are discussed in 
Secs.~\ref{SED} and~\ref{models}. Table~\ref{fwhm} lists the FWHM 
measured for the lines detected in the spectrum.  

\begin{table}
\begin{center}
\caption{FWHM, in \kms, of the detected lines in Mrk\,766} \label{fwhm}
\begin{tabular}{lccc}
\hline \hline
Line & FWHM & Line & FWHM \\
\hline
\lbrack Ca~{\sc viii}\rbrack~2.31$\mu$m& $\leq$360 & He~{\sc ii} (n) ~1.16$\mu$m            & $\leq$360 \\
H$_{2}$ ~2.223$\mu$m                   & $\leq$360  & He~{\sc ii} (b) ~1.16$\mu$m            & 1220 \\
Br$\gamma$ (n)                         & $\leq$360  & \lbrack P~{\sc ii}\rbrack ~1.14$\mu$m  & $\leq$360 \\
Br$\gamma$ (b)                         & 1600       & O~{\sc i} ~1.128$\mu$m                  & 1270 \\
H$_{2}$ ~2.212                         & $\leq$360  & Pa$\gamma$ (n) ~1.093$\mu$m           & $\leq$360 \\
He~{\sc i} (n) ~2.058$\mu$m             & $\leq$360 & Pa$\gamma$ (b) ~1.093$\mu$m           & 1720 \\
He~{\sc i} (b) ~2.058$\mu$m             & 1890      & He~{\sc i} ~1.083$\mu$m                 & 780 \\
H$_{2}$ ~2.033$\mu$m                   & $\leq$360  & \lbrack S~{\sc ii}\rbrack ~1.037$\mu$m  & $\leq$360 \\
\lbrack Si~{\sc vi}\rbrack ~1.96$\mu$m & 390       & \lbrack S~{\sc ii}\rbrack ~1.033$\mu$m  & $\leq$360 \\
H$_{2}$ ~1.957$\mu$m                   & $\leq$360  & \lbrack S~{\sc ii}\rbrack ~1.032$\mu$m  & $\leq$360 \\
\lbrack Fe~{\sc ii}\rbrack ~1.63$\mu$m & $\leq$360 & \lbrack S~{\sc ii}\rbrack ~1.028$\mu$m  & $\leq$360 \\
\lbrack Si~{\sc x}\rbrack ~1.43$\mu$m  & 430       & He~{\sc ii} (n) ~1.012$\mu$m            & $\leq$360 \\
Pa$\beta$ (n) ~1.281$\mu$m             & $\leq$360  & He~{\sc ii} (b) ~1.012$\mu$m            & 1250 \\
Pa$\beta$ (b) ~1.281$\mu$m             & 1740       & Pa$\delta$ (n) ~1.004$\mu$m            & $\leq$360 \\
\lbrack Fe~{\sc ii}\rbrack ~1.257$\mu$m & 370       & Pa$\delta$ (b) ~1.004$\mu$m            & 1700 \\
\lbrack S~{\sc ix}\rbrack ~1.252$\mu$m  & 600       & \lbrack S~{\sc viii}\rbrack~0.991$\mu$m & $\leq$360 \\
\lbrack P~{\sc ii}\rbrack ~1.18$\mu$m  & $\leq$360 & \lbrack Ca~{\sc i}\rbrack ~0.985$\mu$m  & $\leq$360 \\
                 ...                    &   ...     & \lbrack S~{\sc iii}\rbrack ~0.953$\mu$m & $\leq$360 \\
\hline
\end{tabular}
\end{center}
\end{table}

The values of FWHM found in Mrk~766 indicate 
that all NLR lines but [S\,{\sc ix}],
[Si\,{\sc vi}] and [Si\,{\sc x}] are spectroscopically
unresolved or in the limit of the spectral resolution.
This agrees with the FWHM of the optical lines measured
by GDP96. They found that H$\beta$, [O\,{\sc iii}], H$\alpha$, 
[N\,{\sc ii}], [S\,{\sc ii}] and [S\,{\sc iii}] had
widths of less than 250\,\kms, well below our spectral
resolution of 360\,\kms. \citet{vvg01}
also reported a FWHM of 330\,\kms\ for [O\,{\sc iii}]
$\lambda$5007. This supports the existence of two
distinct regions within the NLR of
Mrk\,766: one in which low and medium ionization
lines are formed (the classical NLR), and another where the
bulk of coronal lines (the so-called coronal line region, CLR)
are emitted. Note, however, that no asymmetries are 
observed in the line profiles of the NIR coronal lines, consistent
also with the observations of GDP96. This contrast to what
is usually found in high-ionization 
lines in AGN \citep{pen84, eaw97, rod04}. The lack of asymmetries
suggest that the coronal lines are weakly related to 
outflowing material or strong shocks, as is usually claimed 
\citep{eaw97}. They should
mostly originate from interaction between matter and radiation
from the central source. This issue will be discussed in
Sec.~\ref{models}.

In addition to the permitted and forbidden lines, the NIR spectrum of 
Mrk\,766 also displays conspicous molecular H$_{2}$ lines, 
particularly in the K-band. Analysis of this emission is presented 
in \citet{rod04a}, where it is concluded from the analysis
of molecular line ratios and the vibrational and rotational
temperature that thermal mechanisms drive the excitation of 
the H$_{2}$  gas. Moreover, X-rays from the central source
can account for nearly 70\% of the observed emission. The
remaining 30\% was attributted to the starburts component,
mostly via shock heating by supernova remnants. The
molecular spectrum will no be further discussed throught
this text.

\section{Modeling of the broad-band Spectral energy distributions} \label{SED}

In order to model the SED of Mrk\,766, some indication on
which type of clouds may be present in the nuclear region 
is required. A comparative analysis of the observed continuum
of Mrk\,766 to that of other AGN provides a first 
approach to this problem.

\begin{figure}
\includegraphics[width=88mm]{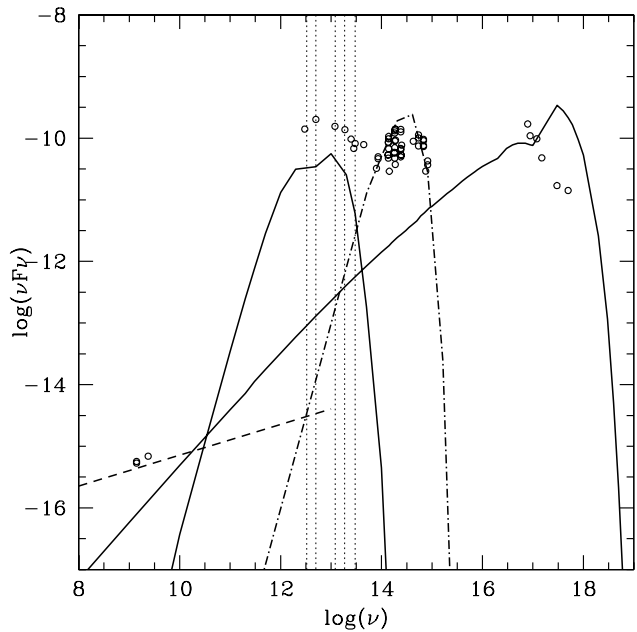}
\caption {Comparison of the observed spectral energy distribution  of 
 Mrk 766 (open circles) and the model  of Ark 564.
The curves correspond
to the fit  to the Ark 564 continuum obtained by detailed
modeling \citep{crv03}. 
The dash-dotted line represents the old stellar population
background. 
The vertical dotted lines refer to the IR wavelengths 
at 10, 16, 25, 60 and 100 \mum. \label{sednls1}
}

\end{figure}

The observed continuum spectra  
of  Mrk~766 (open circles) is plotted in Fig.~\ref{sednls1}. As a guide, 
the energy distribution obtained
in the modeling of Ark\,546 (Paper~I) is also plotted. 
The vertical dotted lines refer to wavelengths in the IR at 10, 16, 25,
60, and 100 \mum. 

 The data come from the NED\footnote{Nasa Extragalactic Database, 
http://nedwwww.ipac.caltech.edu}.
Observations in the optical-UV range
 \citep{gio90, GB96, gio90, dv91} 
 have been corrected for internal reddening (E(B-V)=0.5).
The other data refer to observations at lower frequencies 
 \citep{bw81, DC78, mos90, bwh95, mai95, rlr82, spi95, al03}.
Actually, Alonso-Herrero et al. claim that
the nuclear fluxes at wavelengths greater  than approximately 2~\mum\
will not be significantly affected by moderate amounts of extinction.
Moreover,  the view of the torus through a fragmented screen of dust
 will dominate the flux in the near-IR and  simulate
 a low level  extinction.
No data  appear in the UV where the correction could be significant.

The X-ray data, taken from \citet{mas03} using {\it XMM} observations, have 
been corrected for absorption using the coefficients given by \citep{zom90}
adopting a column density N$_{\rm H}$ = 10$^{21}$ cm$^{-2}$.

In Fig.~\ref{sednls1}, the Mrk\,766 SED 
 is compared to the continuum spectrum obtained from
the detailed modeling of Ark\,564 (Paper I). Recall that the continuum 
which results from modeling
represents reprocessed radiation (free-free and free-bound radiation)
from  clouds reached by the flux from the central source and heated
by the shocks. The main reason for
this comparison is the fact that, overall, both objects have similar 
spectral properties \citep{pag99}. For example,  they do not 
show evidence for the
Big Blue Bump between the ultraviolet and the soft X-ray region.
Moreover, they show a low ratio between the ultraviolet and the 
soft X-ray fluxes, implying a flat optical -- X-ray energy distribution,
as easily seen in Figure~\ref{sednls1}. In fact, for
Mrk\,766 and Ark\,564, $\alpha_{ox}$  
equals to -1.17 (this paper) and -1.08 \citep{rom04}, respectively, indeed flatter than the
mean value of -1.63 for low-redshift radio-quiet objects 
\citep{yua98}. It suggest that, in a first approach, 
the ionizing sources of Ark~564 and Mrk~766 are very similar.

Note, however, that the  flux measurement in the X-rays 
region of Mrk~766  at the highest frequencies are lower
than that for Ark\,564. This  may indicate  that  shock 
velocities as high as found
in the NLR  of the later (V$\geq$1000 \kms) are likely not 
to be present in the former.  The data 
do not exclude a  contribution 
to the X-rays  coming from clouds 
with \Vs\ of about 500 \kms. These type of clouds
are found to be present in Seyfert 2 galaxies \citep{cvp04}.
Because the general behavior of the
Mrk\,766 continuum is different from those observed 
from Sy 2, we expect that if these intermediate
velocity clouds are present in the NLR of Mrk\,766, 
the fraction of the observed emission coming from 
these clouds should differs from the fraction in Seyfert 2.

Another important point is that the flux in the IR region
is higher  for Mrk\,766 than for Ark 564,
which could indicate that the dust-to-gas ratio, {\it d/g}, is higher.
Notice  that {\it d/g} depends on the ratio between
the dust re-radiation  flux intensity  and the bremsstrahlung  
in the optical - NIR range \citep{cvp04}.
As discussed in the following sections, observations
constraints that ratio, and the requirement for high {\it d/g}
becomes less strong.

 In summary, from the comparison of the SED of Mrk~766 and 
Ark~564, we conclude that both objects share a similar continuum
energy distribution properties. It means that a safe starting point to
model the line and continuum emission of the former is to assume
a similar set of clouds successfully used in the modeling of the latter
(Paper I)

\section{Modeling the line spectra} \label{models}

\begin{table*}
\caption{Percent contribution of each model to the multi-cloud model MC} \label{tab:mod}
\begin{tabular}{lllllllllllllllll}\\ \hline
\ & m1&m2&m3&m4&m6&m7&m8&m9&m10&m11&m12&        m15&m16\\ \hline\\
\  \Vs$^1$ &100&100&100&100&150& 150&300&500&500&500&500&          
500&100\\
\  \n0$^2$ &100&100&100&100& 100&600&100&100&100&700&700&         300&1000\\
\  \Fh$^3$
&1.(9)&1.(10)&1.(11)&7.(11)&1.(12)&3.(12)&-&-&5.(12)&-&1.(11)&        1.(11)&-\\
\ D$^4$  &1.&3.&3.&3.&0.01&0.003&3.&3.&3.&3.&1.&             1.&1.\\
\ \Hb $^6$ &0.014&0.1&0.46&0.61&7.4(-3)&0.37
&1.8(-4)&1.9(-3)&24.9&0.015&8.7(3)&678.7&0.66\\
\ d/g$^5$ &10.& 10.&1. & 1.  & 1.&1.&10.  &1.&10.&5. &1.  &10.&0.03 \\
\ wr            & 5.5(-3) &5.5(-3) &2.8(-3)&2.8.(-3)& 8.3(-3)&5.5(-3)
&0.83& 0.083   &1.4(-4) &0.028   &2.5(-6)&1.4(-5) &0.028  \\
\ [Ne{\sc v}] 3425    & -  &-   &-   & 19.   &   - & 50.     &3.1 &-   & 23.  &-    &-    
&-    & - \\
\ [O{\sc ii}] 3727+    &-   & 9.3& 11.& -     &-    &-        & 46.& 5.2&-  &3.8  &-    
&3.4  &16. \\
\ [Fe{\sc vii}] 3758  &-   &-   &-   &  33.  &-    &  8.5    & 10.& -  & 41.  &5.1  &-  
&-     &- \\
\ [Ne{\sc iii}] 3869+ &-  &-   & 9.4&14.    &-     &-        & 7.7& -  & 67.  &-     &-   
& -   &- \\
\ [S{\sc ii}] 4070    & - &9.2 &41.  &-     &-     &-        &4.9 &-   &-    & 17. &18.   
&4.6 & -  \\
\ [O{\sc iii}] 4363   &-  &-   & 8.2& 50.   &-     &  -      &8.9 &-   & 21. & 5.0 &-     
&-   & -\\
\ He{\sc ii} 4686     & - &-   & 3.5&12.    &-     & 24.     & -  & -  & 43.  &-    &6.5  
&7.4  &-   \\
\ [Ar{\sc iv}] 4740   &-  &-   &-   & 49.  &-     &   -     &  - &-   &44.  &- &-     &-  
 &-   \\
\ [O{\sc iii}] 5007+  &-  & -  &  19.& 36.  &-     & -        & -  &-  & 17. & - &13.    &
7.2 &-    \\
\ He{\sc i} 5876      &-  & -  &-    &-     &-     &-         &-   &  - &-   & -  & 6.1  
&  -  &90.  \\
\ [Fe{\sc vii}] 6086  &-  &-    &-   & 32.  &-     &   9.2    &6.6 &  - & 47.&3.4&  -   
&-    &-  \\
\ [O{\sc i}] 6300     & 4.5&27. & 37.&-      &-    &-        &   -  &5.  &- &23. &-     
&-    &-  \\
\ [Fe{\sc x}] 6375    &-  &-    &   -& 20.   &-    &40.      &  4.4 &3.1 &23.& 9.1 &-    
&-    &-  \\
\ [N{\sc ii}] 6548+  & 4.0&  24. &23.&-      &-    &-        & 8.3  &3.8 &-  & 7.6& 18. 
&11.  & - \\
\ [S{\sc ii}] 6717   & 3.7&  25.& 57.&-      &-    &   -     & 3.9  &3.5 &-  & 4.  &   - 
&-    &-  \\
\ [S{\sc ii}] 6731   & 3. &  21.& 54.&-      &-    &-        & 4.4  &4.3 &-  & 7.3  & -  
&-    & - \\
\ [Ar{\sc iii}] 7135 &-   &-    & 16.&-      &-    &-        &      &    &- &-     & 54.
&23.  & - \\
\ [O{\sc ii}] 7320+  & -  &-    &-   &-      &-    & -       & 37.  &3.6 &-  & 22.  &3.  
&-    &28. \\
\ [S{\sc iii}] 0.95  & -  &3.9  &  14.&-     &-    & -       &   -  &  - &- &-     & 51. &
26. & -  \\
\ [S{\sc viii}] 0.99 &-   &-    & -  & 20.   & -    & 51.    &   -  & -  &22.  & -   &-   
& -   & -   \\
\ [S{\sc ii}] 1.028  & -  &9.0  &39. & -     &-    & -       &4.9   & -  &-  & 17. & 18. &
 4.6&  - \\
\ He{\sc i} 1.08     &-   &-    &-   &-      &-    & -       &-     &-   & - &-     &19. 
& 6.2 &71.\\
\ [S{\sc ix}] 1.25   &-   &-    &-   & 22.   &  -  &  38.    &3.9   &-  &24.&7.1   &-   
&-    &-   \\
\ [Si{\sc x}] 1.43   &-   &-    &-   & 20.   & 6.6 & 38.     &  -   &-   & 26.&6.   &-   
&-    &-   \\
\ [Si{\sc vi}] 1.96  &-   &-    &-   &29.    &-    &11.      &   -  &-   & 52.&-    &-   
&-    &-  \\
\  He{\sc i} 2.06    & 16.& 39. &10. & -     &-    &-        &-     &-   &-  &-    &-   
&33.  &-  \\
\ [Mg{\sc viii}] 3.03 &-  &-    &-   &75.    &4.4  &-        & 11.  &-   &-  &7.2  &-   
&-    &-  \\
\ [Si{\sc ix}] 3.93   &-  &-    &-   & 68.   &9.9  &-        & 6.5  &3.7 &-  & 12. &-   
&-   &-   \\
\hline\\

\end{tabular}

$^1$ in \kms ;
$^2$ in \cm3 ;
$^3$ in photons cm$^{-2}$ s$^{-1}$ eV$^{-1}$ at 1 Ryd ;
$^4$ in 10$^{19}$ cm ;
$^5$ in 10$^{-15}$ ;
$^6$ in \erg

\end{table*}

\begin{table*}
\caption{Comparison of observational data with model results. The indices MC and MCS
identify the solution obtained with the two different continua used in the modeling. 
For model MC, the input continuum is the one from the best matching model 
($\alpha_{UV}$ = -1.5 and $\alpha_X$ = -0.4). For MCS, the input continuum is the
one derived by \citet{cs03}. Se text for further details.} \label{tab:comp}
\begin{tabular}{lllllllll}\\ \hline
\     & (I$_{\lambda}/I_{[OIII]})_{obs}$ &
(I$_{\lambda}/I_{[OIII]})_{MC}$&  R([O\,{\sc iii}])$_{MC}$& R(\Hb)$_{MC}$& 
R([O\,{\sc iii}])$_{MCS}$& R(\Hb)$_{MCS}$  \\
\hline
\ [Ne{\sc v}] 3425+ &22.5     &20.6  & 0.91  &1.23&1.64&4.8 \\
\ [O{\sc ii}] 3727+& 13.18    & 7.74  & 0.59 &0.79&1.96&5.7\\
\ [Fe{\sc vii}] 3758 & 2.37  & 2.99  & 1.27& 1.7&2.5&7.4  \\
\ [Ne{\sc iii}] 3869+ & 15.8 & 13.7 &  0.87 &1.17&0.5&1.44\\
\ [S{\sc ii}] 4070+ & 1.5   &  1.3&0.86&1.16&0.85&2.5\\
\ [O{\sc iii}] 4363 & 2.05& 1.81 &0.9  &1.2&0.84&2.5 \\
\ He{\sc ii} 4686 & 4.34  & 5.2  & 1.2   &1.6&3.08&9.12 \\
\ [Ar{\sc iv}] 4740+ & 0.74& 2.1  &2.9  &3.9&2.4&2.4\\
\ [O{\sc iii}] 5007+& 100.& 100. &1. &1.34&1.&3.\\
\ He{\sc i} 5876 & 8.14  & 37.2& 4.6  &6.14&0.3&0.9 \\
\ [Fe{\sc vii}] 6086 & 1.42& 3.43& 2.4   &3.25&5.4&16.\\
\ [O{\sc i}] 6300+ & 2.4   &2.37&0.99 &1.33&1.28&3.8\\
\ [Fe{\sc x}] 6375 & 0.27 & 1.65& 6.1  &8.26&4.7&14.\\
\ [N{\sc ii}] 6548+ & 13.0& 6.6  &0.51&0.68&0.6&1.65\\
\ [S{\sc ii}] 6717 & 2.45 & 4.7 &1.9  & 2.6&0.68&2.01 \\
\ [S{\sc ii}] 6731 & 3.15 & 4.9 & 1.57 &2.1&0.7&2.1 \\
\ [Ar{\sc iii}] 7135 & 1.89 & 1.9  &1.0  &1.35&0.59&1.75\\
\ [O{\sc ii}] 7320+& 1.11  & 1.1 & 1.05 &1.41&2.9&2.02\\
\ [S{\sc iii}] 0.95+& 12.16 & 10. &0.82&1.1&0.5&1.5 \\
\ [S{\sc viii}] 0.99& 1.1  &0.98 & 0.89&1.12&0.7&2.02\\
\ [S{\sc ii}] 1.28 & 1.04 & 0.9   &0.85 &1.13&1.07&3.2 \\
\ He{\sc i} 1.08 &21.2   & 39.5&1.86   &2.5 &0.18&0.53\\
\ [S{\sc ix}] 1.25 & 0.79 & 0.26& 0.33 &0.45&0.22&0.66\\
\ [Si{\sc x}] 1.43 & 0.53 &0.57  &1.07&1.44&0.52&1.53\\
\ [Si{\sc vi}] 1.96 & 1.1 &1.6   &1.5  &2.0&3.5&10.5 \\
\  He{\sc i} 2.06& 0.57   & 0.13 & 0.23 &0.31&0.62&1.83\\
\ [Mg{\sc viii}] 3.027&1.74 &0.61 & 0.35 & 0.48&0.63&1.85\\
\ [Si{\sc ix}] 3.93& 1.10  &1.77 &1.6 &2.15&2.1&6.3\\
\hline

\end{tabular}

\end{table*}

\begin{table*}
\caption{Observed and predicted UV emission lines for the model MC.} \label{tab:uv}

\begin{tabular}{lccccccccc}\\ \hline

Line & F$_{obs}^1$ & F$_{obs}$/\Hb & (I$_{\lambda}$/\Hb)$_{obs}^2$ &
(I$_{\lambda}$/\Hb)$_{MC}$&  R(H$\beta$)$_{MC}$ \\
\hline\\
\ Ly$\alpha$  1214 & 29.78$\pm$0.32  &   0.19  &   5.46 & 29. &  5.4   \\
\ N\,{\sc v} 1240         &       2.10$\pm$0.17   &    0.013  &   0.31&1.0  & 3.23
  \\
\ C\,{\sc ii} 1335        &        1.15$\pm$0.19  &     0.007 &    0.10&- &-\\
\ Si\,{\sc iv}+O\,{\sc iv}] 1400 &  4.31$\pm$1.25   &   0.027  &   0.30&0.75 &   2.5      
       \\
\ N\,{\sc iv}] 1486      &       1.81$\pm$0.82  &     0.012 &    0.11&0.42  &  3.8 \\
\ C\,{\sc iv} 1549      &         17.63$\pm$0.94  &   0.112  &   0.88 &6.1 & 6.9  \\
\ He\,{\sc ii} 1640      &         5.13$\pm$0.52   &    0.033 &    0.24&1.1  &  
4.5   \\
\ Si\,{\sc iii}] 1892   &           2.05$\pm$0.87 &     0.013 &    0.11&0.13 & 1.17\\
\ C\,{\sc iii}] 1908     &          11.75$\pm$1.38  &   0.075  &   0.66&0.38   &
 0.57\\
\ C\,{\sc ii}] 2327     &         4.07$\pm$1.68    &    0.026 &    0.23 &- &-\\
\ [Ne\,{\sc iv}] 2423    &     4.45$\pm$0.70    &    0.028  &   0.17&0.15  &  0.9 \\
\ Mg\,{\sc ii} 2797     &       26.26$\pm$1.41  &    0.17   &   0.51&0.24 &  0.47\\
\  \Hb         &          157.3$\pm$5.60 &     1.   &     1.  &1.  & 1. \\
\hline

\end{tabular}

$^1$ in 10$^{-15}$ \erg ;
$^2$ corrected for E(B-V)=0.5

\end{table*}

By definition, the FWHM of the broad components of the 
permitted lines in NLS1s are $\leq$ 2000 \kms\ \citep{op85}. 
Because the broad lines are narrower than those
of classical Seyfert~1  galaxies,  it is
often difficult to unravel the narrow and broad components of
permitted lines.  This usually leads to an incorrect determination
of the flux contribution of the hydrogen lines emitted by the NLR, 
which in turn could induce large errors when comparing 
the observed and calculated line ratios.  In order to
circumvent this problem, we chose to  normalize  the line 
intensities relative to that of 
[O\,{\sc iii}] $\lambda$5007+4959 emission line flux as is
shown in Table~\ref{derfluxes}.

The criteria adopted to  model  Mrk\,766 follow the
same line of reasoning adopted for Ark~564 (Paper I)
and is summarized as follows: (1)
explain the emission-line spectrum  and  fit  the continuum SED; (2)
the velocities should be in agreement with the FWHM of the line profiles;
(3) the  ratio of calculated and observed  lines relative to \Hb,
R(H$\beta$)$_{MC}$,  must be larger than unity, considering
that the observed \Hb ~flux contains 
also the contribution of the BLR.

A  grid of models calculated  for Ark\,564 (Table~2 of Paper~I)
is rather complete,  covering  a large range
of conditions suitable to describe the NLR of Seyfert galaxies.
Moreover, recently, another 
specific  grid of models was used to discuss the infrared 
continuum of Seyfert galaxies \citep{cvp04}.
All these 
results will be used as a starting point
to explain the emission-line spectrum of Mrk\,766.

In the previous section, we saw that the continuum SED of
Mrk\,766 and Ark\,564 are similar. It
means that a starting point in explaining the emission-line
spectrum of Mrk\,766 should be the use of a multi-cloud model
similar to that adopted for Ark\,564.  However, 
as suggested by the FWHM of the observed line profiles 
in  Mrk\,766,
the high \Vs ~clouds found in Ark\,564 (Paper I, models m13 and
m14 with \Vs=1000 \kms and \Vs=1500 \kms, respectively) 
are not expected to appreciably contribute to the 
observed spectrum.
Thus, a first multi-cloud
model is obtained for Mrk\,766 adopting the same single-clouds used 
in the Ark\,564 but with a different set of  relative weights,
which does not include models m13 and m14. 
 
In addition, a recent analysis of the optical - infrared continuum spectra
of AGN, \citet{cvp04}
showed that low \Vs ~(100 \kms) - high density (1000 \cm3) clouds are present
in all types of Seyfert galaxies as well as clouds with
intermediate velocity and density (\Vs\ = 500 \kms\ and 
\n0\ = 300 s$^{-3}$).  Since the observed continuum of Mrk\,766
can not {\it a priori} exclude these type of clouds from
the NLR modeling, a more consistent
approach can be  obtained 
by including them (models m15 and m16, Table~\ref{tab:mod}). 
The relative weights of
the single clouds were changed in order to obtain  a
better fit to the observed emission-line spectrum. 
The physical properties of the single-cloud models, their \Hb ~absolute flux  
and the relative weights corresponding to the selected multi-cloud
model (MC) are listed in the first rows of Table~\ref{tab:mod} (top).
Table~\ref{tab:mod} also lists the contribution
(in percent) of every single-cloud model to each  emission-line.
The (+) symbol near the line wavelengths indicates that the 
doublet intensities are summed up.
We have removed from Table~\ref{tab:mod} the contributions smaller than
3\%  in order to have a clear view of the results.
The comparison between the observed and modeled emission-line ratios 
is made in Table~\ref{tab:comp} and~\ref{tab:uv}.

\begin{table}
\caption{Percentage contribution of the individual clouds to the 
multi-cloud model MCS, calculated with steep spectral indices.}\label{tab:steep}
\begin{tabular}{lccccccllll}\\ \hline
\ &m6&m7&m8&m9&m10&m13\\ \hline
\  \Vs $^a$ &150& 150&300&500&500&1000\\
\  \n0 $^b$ & 100&600&100&100&100&1000\\
\  \Fh $^c$ &1(12)&3(12)&-&1(10)&5(12)&5(12)\\
\ D$^d$  &.01&.003&3.&3.&3.&.1\\
\  \Hb (MCS)$^e$ &7.(-3)&0.65&1.9(-4)&5.72&62.4&3.5\\
\ wr(MCS) &0.21&8(-4)&0.79&1(-5)&2(-5)&1.3(-4)\\
\ [NeV]3425 & 84.&-&11.&-&-&-\\
\ [OII]3727&-&81.&17.&-&-\\
\ [FeVII]3758 &63.&-&32.&-\\
\ [NeIII]3869+ &4.&30.&27.&-&28.&4.7\\
\ [SII]4070 &-&-& 21.&-&-&73.\\
\ [OIII]4363  &-&26.&59.&-&7.5&-\\
\ HeII 4686 &50.&12.&-&-&36.&-\\
\ [ArIV]4740 &-&70.&15.&-&14.&-\\
\ [OIII]5007+&5.9&43.&14.&-&37.&-\\
\ HeI 5876 &4.4&4.5&8.3&-&9.7&75.\\
\ [FeVII]6086 &75.&4.4&19.&-&-&-\\
\ [OI]6300 &-&-&15.&23.&-&62.\\
\ [FeX]6375 &10.&-&36.&-&-&53.\\
\ [NII]6548+ &-&-&43.&33.&-&23.\\
\ [SII]6717 &-&-&74.&4.&8.4&12.\\
\ [SII]6731 &-&-&63.&4.5&11.&20.\\
\ [ArIII]7135 &-&9.5&-&31.&58. &-\\
\ [OII]7320+ &-&-&77.&8.7&-&14.\\
\ [SIII]0.95& -&-&17.&28.&51.&-\\
\ [SVIII]0.99&64.& - & 23.&-&-&13.\\
\ [SII].028&-&-&22.&-&-&73.\\
\ HeI 1.08 &-&22.&8.9&-&18.&47.\\
\ [SIX]1.25 &6.4&-&35.&-&-&58.\\
\ [SiX]1.43 &-&-&24.&-&-&75.\\
\ [SiVI]1.96 &81.&9.1&6.4&-&-&-\\
\  HeI 2.06&-&-&-&98.&-&-\\
\ [MgVIII]3.03&54.&-&40.&-&-&5.7\\
\ [SiIX]3.93 &17.&-&31.&-&-&52.\\
\hline

\end{tabular}

$^a$ in \kms ;
$^b$ in \cm3 ;
$^c$ in photons cm$^{-2}$ s$^{-1}$ eV$^{-1}$ at 1 Ryd;
$^d$ in 10$^{19}$ cm;
$^e$ in \erg .

\end{table}

 The multi-cloud model MC was calculated assuming 
that the radiation  flux from the active centre is  a power-law
 (F $\propto$ $\nu^{\alpha}$) of intensity  \Fh at 1 Rydberg
(in units of photons cm$^{-2}$ s$^{-1}$ eV$^{-1}$). The spectral
indices $\alpha_{UV}$ = -1.5 (between 13.6 eV $\leq$ E  $\leq$ 260 eV) 
and $\alpha_X$ = -0.4 (for E$>$ 260 eV) were obtained from the best
matching model since the spectrum of Mrk~766 is so 
rich in emission lines and continuum observations 
that it allows us to constrain the input continuum that produces them.
This approach is preferred over the one that directly takes the observed
far UV and X-ray data as input SED because the continuum that
the NLR sees may differ from that emitted by active nucleus.
In fact, other authors (e.g. \citealt{sak03, pag01, vf03, cs03}) 
have found  spectral indices  steeper than those used here for the 
ionizing radiation.  In order to test this hypothesis, we have calculated
 single-cloud models assuming  an ionizing spectrum with 
a power law index $\alpha_{UV}$=-2.4  and $\alpha_{X}$ = -1.7,
which correspond to the indices reported by  \citet{cs03}. 
These new models were obtained with the same 
\Vs ~and \n0 as shown in Table~\ref{tab:steep} and were used to built 
the multi-cloud model labeled MCS.

Although a reasonable fit to the data can be obtained
with the cloud models listed in Table~\ref{tab:steep},  
we found that a high fraction of the line fluxes, corresponding 
to both high and low -ionization lines, are originated
from high velocity clouds (\Vs=1000 \kms, model m13 of Paper I).
Such type of clouds are excluded because
the measured FWHM of the line profiles in the nuclear spectrum
of Mrk\,766 rules out the existence of gas motion with 
velocities above 500\,\kms\ within the NLR. We therefore adopt  model MC 
discussed above to obtain the SED of the continuum and to 
compare its prediction to the observations. 
Notice that the observed soft X-ray index is -1.77, while for the
modeling we adopt -0.4. The discrepancy between both indices
suggests that the NLR clouds see a harder, different continuum,
than that we detect. This same conclusion had previously been
reached by several authors \citep{bin89, bfp93, kfb97}  
in studying the physical conditions
of the BLR using photoionization models. Moreover, in our models,
the observed soft X-ray contiuum is due to 
bremsstrahlung orginating in the clouds that re-process the 
central ionizing radiation.

Before concluding this section, it is important to  
comment on the values listed in Table~\ref{tab:comp},
where the ratio between the calculated and observed line
ratios relative to [O\,{\sc iii}], R([O\,{\sc iii}])$_{\rm MC}$, and to 
\Hb, R(\Hb)$_{\rm MC}$, are listed.

The  calculated emission-line ratios for most lines except
[Mg\,{\sc viii}] and [S\,{\sc ix}] are,
 within the error bars, good by a factor of 2.
The latter two lines are underestimated by a larger factor.
  
It is found that a large contribution to the bulk of the
line fluxes comes from radiation-dominated clouds. This is
the case, for example, of the NIR coronal lines. In them,
the radiation flux from the active centre strongly contributes
to the ionization of gas, which is heated to high temperatures
(T$_{\rm e}$~$>$ 1.5 10$^5$ K) by the shock. Note that even in 
radiation-dominated clouds the shock is also accounted for. 
On clouds propagating outwards from the AC, it acts on the 
edge opposed to that facing the radiation
flux. The coupled effect of the  shock and the radiation flux 
determine the distribution of the temperature throughout the 
cloud. The temperature in the postshock region, immediately after the
shock front is T$_{\rm e}$ = 1.5$\times$10$^{5}$\,(\Vs/100)$^{2}$ K decreasing  
downstream due to energy losses by free-free, free-bound,  
line emission, and collision heating of dust. The stratification 
of the ions downstream follows the distribution of the temperature
described in \citet{cv91}. Our results suggest that the coronal
lines in Mrk\,766 are formed mainly from the interaction of
matter and radiation from the central source rather than shocks.
A similar result is found from the analysis of optical and
NIR coronal lines in a sample of Seyfert 1 and 2 galaxies by 
Rodr\'{\i}guez-Ardila et al. (2004, in preparation). 
They report, using high-spatial resolution spectroscopy, 
extended coronal emission within the few tens of parsecs from the
active nucleus and points out that radiation from the central source
are, in fact, able to power them. Support for this scenario is
also provided by  \citep{pri04} by means of VLT
NIR observations of Circinus.
Other evidence, however,  supports the
hypothesis that coronal lines are most likely associated to 
outflowing winds\citep{eaw97, rod02b}. 

It is also possible
to have a strong contribution from shock-dominated clouds,
as is, for example, the ratio [O\,{\sc ii}]3727+/[O\,{\sc iii}]5007+. 
The high value of the ratio comes from the fact that 70\% of the  
[O\,{\sc ii}] 3727+ line flux comes from that type of clouds 
(models m8, m9, m11, and m16).

The optical [N\,{\sc ii}]+ lines are underestimated by the 
multi-cloud model MC by about a factor of 2. Since
both R([O\,{\sc iii}])$_{MC}$ and R(\Hb)$_{MC}$ are less than
unity, we conclude that the $N/H$ abundance is
about a factor of 2 higher than solar. 

On the other hand, all Fe lines  are overestimated,
indicating that Fe is depleted, probably  trapped into small grains.
Also, [Ar\,{\sc iv}] 4740+ are overestimated, perhaps due to a too high
$Ar/H$ relative abundance. 
We notice that in order to improve the results for the Fe lines, both
model MC and MCS require that $Fe/H$ be
depleted  by  a factor of $\sim$3. Furthermore,
ratios of calculated to observed line ratios (relative to \Hb),
indicate an  average  contribution of the broad line region to the
observed \Hb ~of about 40\% (see columns 5 and 7 of Table~\ref{tab:comp}). 
In Ark\,564, that contribution 
was found to be lower ($\sim$30\%). 

It is important to note that
photoionization models for the NLR and the extended
emission region of Mrk\,766 were also presented by GP96.
They used two types of ionizing continua: a power-law, with 
spectral indices -1 and -1.5 or a blackbody with
effective temperatures $T_{\rm eff}$=10$^{5}$\,K and
2.2$\times$10$^{5}$\,K. The comparison between the 
calculated and observed emission line ratios of [O\,{\sc iii}]/H$\beta$,
[O\,{\sc ii}]/H$\beta$, [O\,{\sc i}]/H$\alpha$,
[N\,{\sc ii}]/H$\alpha$, [S\,{\sc ii}]/H$\alpha$ and
[S\,{\sc iii}]/H$\alpha$ showed that both the power-law
and blackbody could explain the observed data.
Clearly, the inclusion of emission lines
located in other wavelengths intervals (NIR and UV), as
was done here, gives more consistency to the modeling and
allows a better understanding
of the physical conditions in the NLR of Mrk\,766.

Finally, the observed and calculated UV lines  relative to
\Hb ~are compared in Table~\ref{tab:uv}. The results show that there
is a high  contribution  of model MC, i.e., of the NLR,
to the UV lines, particularly for N\,{\sc v}, N\,{\sc iv}]
C\,{\sc iv} and He\,{\sc ii}.

\section{The continuum SED of Mrk766} \label{modSED}

\begin{figure}
\includegraphics[width=86mm]{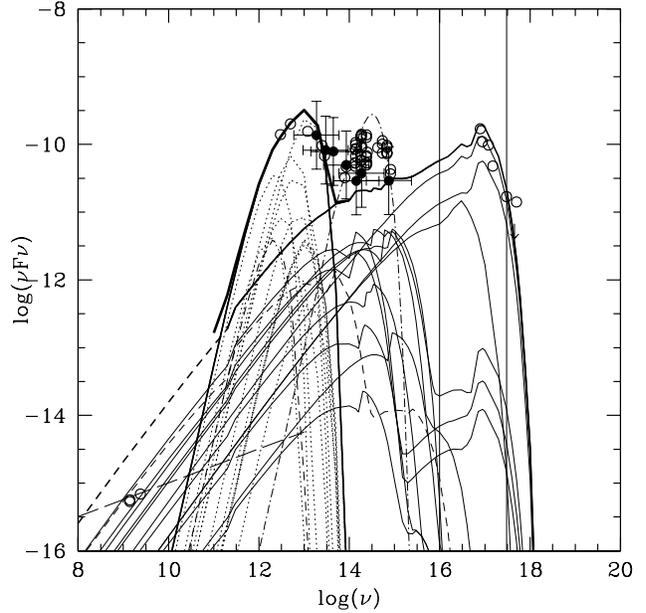}
\caption {The observed and calculated continuum of Mrk\,766. The
data come from NED (open circles) and \citet{al03}. (solid
circles). The thin lines correspond to the results from the single clouds
contributing to the best multi-cloud model MC, with dotted lines
representing dust emission and solid lines  bremsstrahlung emission from
those clouds. The two vertical solid lines define the frequency range
in which the X-ray radiation is absorbed.
The spectrum obtained with the low velocity,
high density single-cloud model m16 is shown by the thin dashed line.
The very thick solid line shows the  continuum corresponding
to the multi-cloud  model MC. \label{fig:SEDmod}}
\end{figure}

\begin{figure}
\includegraphics[width=86mm]{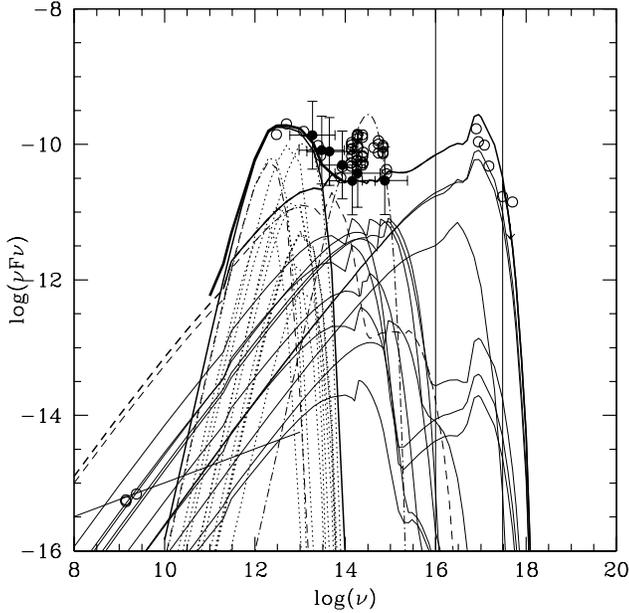}
\caption {The observed and calculated continuum of Mrk 766. The
multi-cloud model was derived assuming a relative weight for
the low velocity, high density single-cloud (model m16) a factor
10 higher than for MC (see text). Same notation  as in Figure~\ref{fig:SEDmod}.
\label{fig:sedm16}}
\end{figure}

In order to  consistently model the  line and continuum
spectra, it is necessary to
cross check one another until a fine tune for both is found.
Therefore, the results presented in the previous section were 
selected after this checking was performed.

The  calculated SED of Mrk\,766 obtained from model MC
is compared to the observations in Fig.~\ref{fig:SEDmod},
where the data taken from NED  (open circles) and from 
\citet{al03} (solid circles) are plotted.
Notice that the errors  for the Alonso-Herrero et al.'s data  are about 30\% .
The thin dotted lines and the thin solid lines represent
the  emission from the single clouds due to
dust  and  bremsstrahlung, respectively.
The thin dashed lines correspond to the results from  model m16
(see \citealt{cvp04}, Sect. 3),  which is included 
in the multi-cloud model MC with other single-cloud models used to
explain the Ark\,564 spectra.
The thick solid line represents the multi-cloud model with separated
curves representing re-radiation by dust and bremsstrahlung, while  
the very thick line represents the sum of these
two curves.
Finally, the dash-dotted line corresponds to the background old star
population.

Another SED,  obtained by adopting a  relative weight
for model m16 higher by a factor of 10 than the value used 
in model MC, is  also shown in Fig~\ref{fig:sedm16}.
However, regarding the corresponding emission-line spectrum,  this 
alternative multi-cloud model leads to very low line intensities 
relative to  \Hb ~(R(H$\beta$)$_{MC}$) and was dropped.

The data in the radio range (long-dashed line)  
show that radiation at these frequencies is due to
synchrotron created by the Fermi mechanism at the shock front.
At  long wavelengths,  bremsstrahlung emission is self-absorbed,
particularly for models with a high density downstream,
which depends on \n0, \B0, and \Vs.
Actually,  the optical thickness becomes larger than unity for
T$\leq$ 10$^4$ K, $\nu$ $\leq$ 10$^{11}$ Hz, for a slab of at least 1.3 pc
downstream of high density clouds (e.g. model m12) because compression
leads to densities higher than 2$\times$10$^5$ \cm3 \citep{ost88}.
Thus,  the contribution to bremsstrahlung in the radio range of
relatively high density clouds   must be negligible.

Overall, the observed and calculated SED are in very good
agreement, as can be seen in Figure~\ref{fig:SEDmod}. This
confirms and strengthens the validity of our approach. The 
largest discrepancy is located in the X-ray region, where
the model overpredicts the observed emission. Here, the rapid 
and strong variability of the X-rays radiation, of up to a 
factor of 2 on a time scale of a few thousand seconds \citep{mas03}, 
clearly indicates that the bulk of this radiation 
is directly emitted by the central source and the BLR. 
However, it is expected that a 
small fraction of the X-ray emission arises 
from high velocity clouds. Indeed, the bremsstrahlung 
calculated by models m11 and m12 overpredict the data.
Notice however that 
a possible absorption of the X-rays by these kinds of clouds is not taken into
account. In fact, absorption must occur firstly because of  the 
large geometrical thickness of the clouds (D=10 pc)  with \Vs=500 \kms\ 
and \n0=700 \cm3, and secondly, because of the high densities downstream
due to compression (n/\n0 $\leq$ 10, depending on \Vs, \n0, and \B0).
Under these conditions, column densities higher than 10$^{23}$ cm$^{-2}$
occur in the clouds producing a strong absorption of the high energy radiation
from 13.6 eV up to 500 eV \citep{zom90}. Clouds with lower values of 
D, \Vs  and \n0 unlikely absorb X-rays.

\section{Discussion and  concluding remarks} \label{fin}

As part of a wide program to understand the mechanisms at work
in the NLR of NLS1 galaxies, we have carried out a detailed
study of the galaxy Mrk\,766. For this purpose, the first 
spectroscopic  observations covering simultaneously the
interval 0.8--4~$\mu$m is presented
along with the measurement of the most conspicuous emission line
fluxes detected in that region. These lines are used as additional constraints 
in order to unveil the physical conditions of the NLR gas. 
The single-cloud  models adopted for modeling are those
previously  used for the analysis of the continuum and
emission-line spectra of Ark\,564.
However, the two high velocity cloud models used for the later object
were removed from the multi-cloud model that fits Mrk\,766
because the X-rays flux is lower when compared
to the continuum emission in other frequency ranges
and the FWHM of the emission lines indicate lower velocities.
On the other hand,  two cloud models suggested by the fit of the 
NIR data of \citet{al03} sample were added. Indeed, 
after  including these single-cloud models, a better fit 
of the line intensities  relative  to [O\,{\sc iii}] and 
\Hb, as well as of the continuum, were be obtained. 

As already found in the analysis of other individual AGN,
shock dominated models have higher relative weights because the
absolute \Hb ~flux is lower.
The shock velocities range between 100 \kms and 500 \kms,
the preshock densities between 100 \cm3 and 1000 \cm3 and the  
radiation fluxes from the active center from 10$^9$ to 5 10$^{12}$  photons
cm$^{-2}$ s$^{-1}$ eV$^{-1}$ at 1 Ryd with  spectral indices of
$\alpha_{UV}$=-1.5 and $\alpha_X$=-0.4. Higher power-law indices were dropped
because they did not lead to satisfactory explanation of the
observed spectra.

The {\it d/g} ratios are between  10$^{-15}$ and 10$^{-14}$ by number
(4$\times$10$^{-5}$ to 4$\times$10$^{-4}$ by mass for silicates)  and reach the
 lowest value in clouds
corresponding to model m16 with {\it d/g} = 3 10$^{-17}$
($\geq$ 10$^{-6}$ by mass).

We consider that R=A$_{\rm V}$/E(B-V) $\sim$ 3.1 \citep{ccm89} and
 A$_{\lambda}$=1.086 $\tau_{\lambda}$ with $\tau_{\lambda}$=Q$_{\lambda}$ $\pi$ \agr
$^2$ N$_{gr}$, 
where Q$_{\lambda}$ is the extinction efficiency, \agr the grain radius,
and N$_{gr}$ is the column density of dust.
N$_{gr}$ = n $\times$ d/g $\times$ $x$, where n is the gas density downstream
which increases with \n0 and \Vs, and $x$ is the geometrical
thickness of the dust slab.
Adopting E(B-V)=0.5, Q$\sim$ 2, n=10$^4$-10$^5$ \cm3, d/g = 10$^{-15}$ - 10$^{-14}$,
$x$ $\sim$ 0.2 - 20 pc in good agreement with the geometrical thickness of the
clouds  given in Table 2.

The important contribution to the NIR SED of Seyfert galaxies
of high density-low velocity clouds,
corresponding to model m16, was demonstrated  by \citet{cvp04},
who also raised  questions  about their location in the galaxies and
their dust-to-gas ratios. With the data already discussed, we are  
now able  to roughly estimate the  distance from
the active center. To this purpose, we use  the [O\,{\sc ii}] 7320+ line, 
which is nicely fitted by the multi-cloud model and shows the highest 
percentage contribution  from  model m16
relative to the other forbidden lines. This  type of
cloud is  also responsible for emission of the  permitted lines
He\,{\sc i} 5876\,\AA\ and He\,{\sc i} 1.083\,$\mu$m. However, since 
the latter two lines include the contribution from the BLR,  
any location of the clouds deduced from these lines may be mistaken.
From Table~\ref{tab:mod}, we see that the larger contribution to 
the [O\,{\sc ii}] 7320+ emission-line comes from  the single-clouds 
m8, m10 and m16. Thus, assuming that the single-clouds are
co-spatial, their distance $r$ from the central source
can be estimated from the expression,

\begin{equation}
F_{\mathrm [OIII]~obs} \times d^{2} = (\Sigma_{i} wr(i) \times H\beta(i) \times ([OII]/H\beta)_{\mathrm calc}(i)) \times r^{2}
\end{equation}

where  {\it i}= 1,2,3, refers to the three types of single clouds,
{\it d} is the distance to Mrk\,766, {\it wr(i)} and \Hb(i) are given 
in Table~\ref{tab:mod}, and the line intensity, relative to \Hb, is equal to  
4.64, 1.5, and 0.04, respectively, for m8, m11 e m16.
 Assuming H$_o$=75 \kms /Mpc,
the distance of Mrk\,766 from Earth is {\it d}= 51.6 Mpc.  
Thus, we obtain an average distance of {\it r} = 160~pc. 
Since usually higher velocity clouds (m8 and m11)
 are closer to the center,
{\it r} is probably a lower limit to the distance of the 
low-velocity clouds (m16). The average distance obtained
for the clouds is also in very good agreement to the
region size of 250\,pc 
covered by the nuclear NIR spectrum employed in this work. 
This gives further consistency
to our modeling an approach to describe the physical 
conditions of the NLR of Mrk\,766.

The high \n0 and  the low dust-to-gas ratio  of the clouds m16 
may indicate that they are  the residual of matter which was
already compressed  by  a previous  blast wave  
propagating  outwards.
The velocity of the blast wave was high enough to destroy
most of the dust grains  by sputtering,
 i.e., \Vs $\geq$ 200 \kms \citep{cvp04}.
Besides compression, another effect of the blast wave is fragmentation,
which  follows in the turbulent regime created by shocks. 
Indeed, in  this model,
because of the relative high density, the postshock
cooling rate is high and the gas recombines at a distance  
$\leq$ 10$^{17}$ cm from the shock front.  Beyond 
this distance the gas temperature  drops to  $<$ 500 K.
Thus, the clouds are radiation-bound and the geometrical 
thickness (D) adopted for model m16 can be regarded as an 
upper limit. Note that the clouds 
corresponding to model m16 could be geometrically thin, 
strengthening the hypothesis of fragmentation.  

In view of the important role of NLS1 in the determination of
metallicities in AGN,  in particular through the  nitrogen lines
(e.g. \citealt{sn02}), the results of the multi-cloud model
suggest that a {\it N/H} higher than solar by a factor  $\geq$ 2
would be in agreement with the  [N\,{\sc ii}]+/[O\,{\sc iii}]+ line 
ratio (Table~\ref{tab:comp}). We also suggest that iron is 
depleted from the gaseous phase because of its emission line
ratios are systematically overpredicted. 

Finally, let us conclude that the coupled effects of
photoionization and shocks allowed us to explain most of the 
features observed in both the line and continuum spectra
of the NLS1 Mrk 766.

\section*{Acknowledgments}
{We are grateful to the referee, Dr. D. Grupe, for many valuable 
comments to improve this manuscript.
This paper is partially supported by the Brazilian funding agencies
FAPESP (00/06695-0) and CNPq(304077/77-1 and 309054/03-6) to ARA and 
SMV. This research has made use of the 
NASA/IPAC Extragalactic Database (NED) which is operated by the 
Jet Propulsion Laboratory, California Institute of Technology, 
under contract with the National Aeronautics and Space 
Administration.


\end{document}